\documentclass[
footinbib,
a4paper,
aps,
prx,
superscriptaddress,
reprint,
twocolumn,
preprintnumbers,
nobalancelastpage,
floatfix,
10pt]{revtex4-2}

\usepackage{xcolor}
\usepackage{graphicx}
 	\graphicspath{ {figuras/} }
\usepackage{dcolumn}
\usepackage{tikz}
\usepackage{subfigure}
\usepackage{float}
\definecolor{cmblue}{rgb}{0.12156862745098039, 0.4666666666666667, 0.7058823529411765}
\definecolor{mygrey}{gray}{0.35}
\definecolor{myblue}{rgb}{0.2,0.2,0.8}
\definecolor{mygreen}{rgb}{0.2,0.8,0.5}
\definecolor{myzard}{cmyk}{0,0,0.05,0}
\definecolor{mywhite}{rgb}{1,1,1}
\definecolor{myred}{rgb}{1,0.,0.3}

\usepackage[utf8]{inputenc}
\usepackage{dsfont}
\usepackage[normalem]{ulem}
\usepackage{bbold}
\usepackage{soul}
\usepackage{slashed}

\usepackage{amsmath, amsthm, amsfonts, amssymb,amstext}
\usepackage{mathptmx}
\usepackage{bm}
\usepackage{xfrac}
\usepackage{mathrsfs}
\usepackage{latexsym}
\usepackage{array}

 \def\ee{\mathord{\rm e}}
 
 \def\ii{\mathord{\rm i}}
\def\min{\mathord{\rm min}}

\def\half{\textstyle\frac{1}{2}}

\newcommand{\diff}{\text{d}}
\def\beq{\begin{equation}}
\def\eeq{\end{equation}}
\def\barray{\begin{eqnarray}}
\def\earray{\end{eqnarray}}

\usepackage{physics}
\usepackage{braket}
\usepackage{tikz}

\usepackage[
colorlinks=true,
citecolor=myblue,
linkcolor=myred]
{hyperref}
\usepackage{cleveref} 
\usepackage{verbatim}

\usetikzlibrary{backgrounds,fit,decorations.pathreplacing,calc}
\usetikzlibrary{quantikz2}
\renewcommand{\ii}{{\rm i}}
\renewcommand{\ee}{{\rm e}}

\begin{document}
\author{P. Viñas}
\affiliation{Instituto de F\'isica Te\'orica UAM-CSIC, Universidad Aut\'onoma de Madrid, Cantoblanco, 28049, Madrid, Spain}

\author{A. Bermudez}
\affiliation{Instituto de F\'isica Te\'orica UAM-CSIC, Universidad Aut\'onoma de Madrid, Cantoblanco, 28049, Madrid, Spain} 


\title{Microscopic parametrizations for gate set tomography  under coloured  noise}

\begin{abstract}
Gate set tomography (GST)  allows for a  self-consistent  characterization of noisy quantum information processors (QIPs). The standard device-agnostic approach  
treats the QIPs as black boxes  that are only constrained by the  laws of physics, attaining full generality at a
considerable  resource cost:   numerous circuits  built from the gate set  must be run in order to amplify each of the gate set parameters.  
In this work, we show that  a microscopic parametrization of   quantum gates under time-correlated  noise on the driving phase, motivated by recent experiments with  trapped-ion gates,  reduces the required resources  enabling  a more efficient version of GST. By making use of the formalism of filter functions over the noise spectral densities,   we discuss the minimal parametrizations of the  gate set that  include  the effect of  finite correlation times and non-Markovian quantum evolutions during  the individual gates.  We compare the estimated gate sets obtained by our method and the standard long-sequence GST, discussing their accuracies in terms of established metrics, as well as showcasing the   advantages of the parametrized approach in terms of the sampling complexity for specific examples.

\end{abstract}

\maketitle

\setcounter{tocdepth}{2}
\begingroup
\hypersetup{linkcolor=black}
\tableofcontents
\endgroup

\section{\bf Introduction}
Unlocking the full potential of quantum information processors (QIPs) is a multifaceted challenge that requires key advances along various directions, both technological and fundamental~\cite{nielsen_chuang_2010}. Among these, the development of efficient techniques for the characterization of undesired interactions -- or {noise} --  in QIPs, 
is of particular relevance, for it  allows to develop improved strategies to counter its undesired effects. This is the case of some schemes~\cite{vandenBerg2023} of quantum error mitigation~\cite{PhysRevLett.119.180509,PRXQuantum.2.040326,RevModPhys.95.045005}, or some strategies~\cite{PhysRevLett.120.050505,
Bonilla_Ataides_2021} for  quantum error correction~\cite{PhysRevA.54.1098,  PhysRevA.54.1098, PhysRevLett.77.793, RevModPhys.87.307} that may allow to find shortcuts towards  the fault-tolerant era.
 In pursuit of this goal, the community working in quantum characterization, verification, and validation (QCVV)  has devised a wide variety of methodologies~\cite{Eisert2020,PRXQuantum.2.010201,Gebhart2023}. Some of these approaches seek to provide a unified metric to  estimate device performance without explicitly detailing the  form of the errors, as exemplified by randomized benchmarking \cite{Emerson_2005,Knill_2008, Dankert_2009, PhysRevLett.106.180504,
PhysRevA.85.042311}. Others strive towards a precise prediction  of the actual faulty operations in the noisy QIPs, belonging to the family of  quantum tomography techniques~\cite{gill2004invitation, Banaszek}.

\begin{figure}
  \centering
  \includegraphics[width=1\columnwidth]{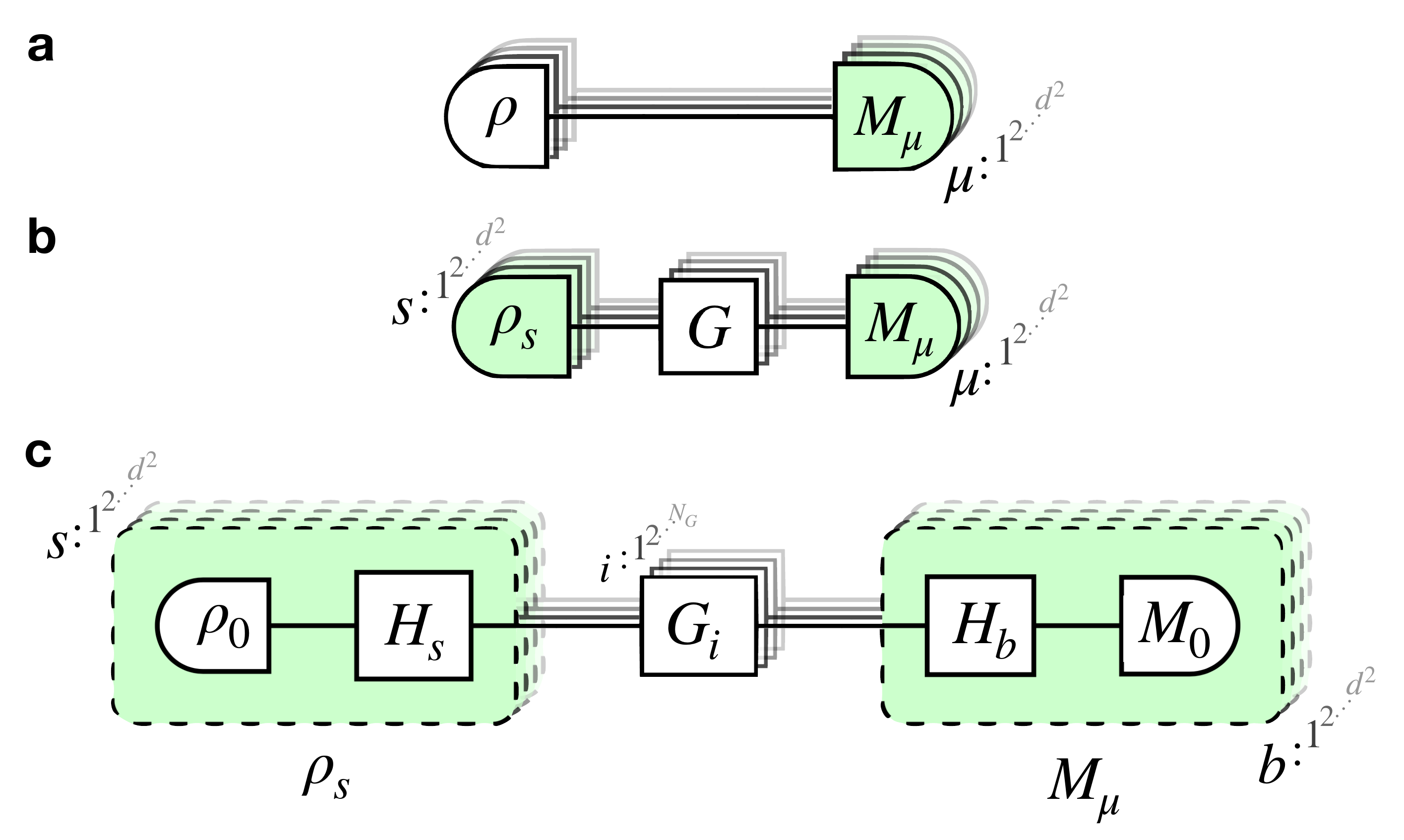}
  \caption{\textbf{Schemes of quantum tomography schemes.} \textbf{a} State tomography. An informationally complete set of $d^2$ predetermined POVM elements  $\{M_\mu\}$ needs to be specified. \textbf{b} For process tomography, in addition to the POVM elements , an IC set of $d^2$ initial states $\{\rho_s\}$ must also be specified. \textbf{c} Gate set tomography. The QPT scheme must be executed for each of the $N_G$ gates at the set. These gates, or combinations thereof (denoted by $H_s$ and $H_b$), act on a single experimentally-accessible initial state $\rho_0$ and POVM $M_0$ to obtain the aforementioned IC sets.}
  \label{fig: tomography_scheme}
\end{figure}

The  simplest protocol that falls in this category is quantum state tomography (QST)~\cite{PhysRevA.40.2847,PhysRevA.55.R1561,Paris2004,Christandl_2012,
Cramer_2010, Blume_Kohout_2010}, which aims to estimate an unknown quantum state  described by a unit-trace positive-semidefinite linear operator in a Hilbert space $\mathcal{H}$ of dimension $d={\rm dim}(\mathcal{H})$, namely $\rho\in\mathsf{D}(\mathcal{H})\subset\mathsf{Pos}(\mathcal{H})$. For discrete-variable systems, QST requires implementing measurements drawn from an {informationally-complete} (IC) set of positive operator-valued measure (POVM) elements $\{M_\mu:\,\mu\in\mathbb{M}\}$, where $M_\mu\in\mathsf{Pos}(\mathcal{H})$ and $\sum_\mu M_\mu=\mathbb{1}_2$~\cite{watrous_2018} (see Fig.~\ref{fig: tomography_scheme}\textcolor{magenta}{a}). These allow for an univocal  estimate $\hat{\rho}$ of the state $\rho$ from the observed probabilities $p_\mu={\rm Tr}\{M_\mu\rho\}$, where we will use hats to differentiate the statistical estimates from the actual objects (see App.~\ref{appendix_2}). 
In quantum process tomography (QPT) \cite{doi:10.1080/09500349708231894,PhysRevLett.78.390,PhysRevA.63.020101,PhysRevA.63.054104,PhysRevA.68.012305}, the focus lies on the estimation of the quantum channel $\rho\mapsto\mathcal{E}(\rho)$, a completely-positive and trace-preserving linear operator  $\mathcal{E}\in\mathsf{C}(\mathcal{H})\subset\mathsf{L}(\mathcal{H})$  that describes a physically-admissible operation on a quantum system transforming input to output  states~\cite{nielsen_chuang_2010,watrous_2018,KRAUS1971311, CHOI1975285}. In addition to the above IC  measurements, an IC set of initial states $\{\rho_{s}:\,s\in \mathbb{S},\,|\mathbb{S}|=d^2\}$ is also required to perform QPT. See Fig.~\ref{fig: tomography_scheme}\textcolor{magenta}{b} for a schematic representation of QPT. Similarly to QST, one can find an univocal estimate of the channel $\hat{\mathcal{E}}$ from the measured probabilities $p_{\mu s}={\rm Tr}\{M_\mu\mathcal{E}(\rho_s)\}$ (see App.~\ref{appendix_2}).

In practice, however, any real experiment will only provide an approximation to above POVM probabilities given a finite number of measurement shots. We consider from now on that each POVM is described by a measurement basis $b$ and a  possible outcome $m_b$, such that $\mu=(b,m_b)\in\mathbb{M}=\mathbb{M}_b\times\mathbb{M}_{m_b}$. This is the case of the Pauli measurements in $N$-qubit systems with $d=2^N$, where the POVM elements can be defined in terms of 
the orthonormal Pauli basis
$
E^{\phantom{\dagger}}_\alpha\in\!\big\{\mathbb{1}_2,\sigma_x,\sigma_y,\sigma_z\big\}^{\!\!\bigotimes^N}\!\!\!\!/\sqrt{d}.$
Each Pauli measurement outcome is binary $m_{b}=\pm 1$, and the specific number of outcomes $N_{b,m_b,s}$ that result from $N_{b,s}$ measurement shots per initial state and measurement basis  lead to a binomial distribution with relative frequencies $f_{\mu,s}=N_{b,m_b,s}/N_{b,s}$.  In the asymptotic limit $N_{b,s}\gg 1$, this tends to a normal distribution $f_{\mu s}\sim{N}(\bar{\mu},\bar{\sigma}^2)$ centered around the aforementioned probabilities  $\bar{\mu}=p_{\mu, s}$ with standard deviation  $\bar{\sigma}=\sqrt{p_{\mu, s}(1-p_{\mu, s})/{N_{b,s}}}$. Hence, there are intrinsic finite-sample errors
that describe the deviations from  the  probability distributions $p_{\mu s}$, which are typically referred to in the literature as quantum projection or shot noise~\cite{PhysRevA.47.3554}.

In analogy to QST,
one can obtain a linear relation between measured relative frequencies and the quantum channel acting as a super-operator~\cite{kitaev2002classical}. For instance, using  the so-called {Pauli Transfer Matrix} (PTM) representation (see Appendix~\ref{appendix_2}), the linear equations can be  straightforwardly  inverted. However, this linear inversion can lead to non-physical estimates of the channel $\hat{\mathcal{E}}\notin\mathsf{C}(\mathcal{H})$ when finite-sample errors are significant. To avoid these problems, {maximum likelihood estimation}  is typically preferred~\cite{rossi2018mathematical}, as the numerical optimization of  a cost function can be supplemented with    constraints to ensure that all estimated channels are indeed physically acceptable~\cite{PhysRevA.63.020101, PhysRevA.63.054104}. One of the advantages of this strategy is that the constrained optimization problem is convex and, thus, a unique estimate $\hat{\mathcal{E}}$ is  guaranteed.


Gate set tomography (GST) (Fig.~\ref{fig: tomography_scheme}\textcolor{magenta}{c}) was introduced to overcome a fundamental shortcoming of QPT~\cite{Nielsen_2021, 
greenbaum2015introduction,
blumekohout2013robust,
PhysRevA.89.052109,Merkel_2013}. QPT, although  used in several QIP platforms~\cite{Childs2001,10.1063/1.1785151,PhysRevLett.91.120402,PhysRevLett.93.080502,PhysRevLett.97.220407,doi:10.1126/science.1177077,PhysRevLett.102.040501,Bialczak2010,PhysRevLett.109.240505},  suffers from a self-consistency issue: it assumes that the IC set of initial states and measurements  are error free. In reality, however,  these states are usually obtained by acting with a set of gates  $\{G_i\}$ on a single fiducial state $\rho_0$, which reduces to $\ket{0}\!\bra{0}$ in the ideal $N=1$ case. Likewise, the measurements are obtained by acting with these gates on   a single binary projective measurement $M_{z,+}=M_0,\,M_{z,-}=\mathbb{1}-M_0$, which reduce to $\ket{0}\!\bra{0}, \ket{1}\!\bra{1}$ in the ideal $N=1$ case. Clearly,  $\{G_i\}$ are implemented similarly to the noisy gate $\hat{\mathcal{E}}$ that  we aim  at characterizing, and will thus contribute to a non-zero  state preparation and measurement (SPAM) error. In the case of single-qubit gates, this error will be of the same of order as the one that afflicts the  gate we want to estimate. It is thus not justified to assume that the SPAM  is ideal and error free. GST  addresses this self-consistency issue by  targeting the full set of noisy elements, including the gates used to attain IC and   as well as the  fiducial   operators in a full ``gate set''  
\beq
\label{eq:gate_set}
\mathcal{G}=\Bigl\{\rho_0, \,\,\big\{G_i:\,i\in \mathbb{G}\,\big\},\,\, M_0\Bigr\}.
\eeq

As emphasised in~\cite{Nielsen2021}, QPT is  not  device-independent in its entirety, as it requires a  prior independent calibration of   the measurement basis and initial states. GST eliminates this requirement and provides a complete self-consistent characterization that treats the whole QIP device as a black box. However, to attain this full generality, it requires at least $N_{GST}=N_G\times4^N(4^N-1)+2^{3N}-1$ parameters for an $N$-qubit system, where $N_G=|\mathbb{G}|$ is the number of gates in the  set~\eqref{eq:gate_set}. Estimating these  parameters thus requires an exponentially-large number  of imperfect fiducial state preparation, followed by a combination of gates from the gate set, and  a final fiducial imperfect measurement. In addition to this exponential scaling, the number of samples required to achieve a target precision $\epsilon$ increases as $N_{b,s}\propto1/\epsilon^2$. Altogether, the resource scaling limits the applicability of a fully-general GST to systems with small qubit numbers. The development of long-sequence GST  is a significant advancement in this regard \cite{Nielsen2021gatesettomography}, which estimates the gate set parameters using data from specific circuits that are built from combining elements of the gate set.  This protocol linearly amplifies gate parameters as the circuit depth $p$ increases, allowing for a Heisenberg-like scaling of the  precision estimates. In particular, in the asymptotic limit, the error  falls with  the inverse of the circuit depth  and not with its square root, as would occur if one only used the resources to increase the number of repeated measurements in circuits that only involve a single gate in addition to the SPAM ones. 


The full generality of GST is convenient  to validate and verify the functioning of a  certain QIP device, regardless of having any specific microscopic knowledge, making it useful for  external users. On the other hand, it can also help developers to calibrate their devices and guide hardware progress in the current noisy intermediate scale quantum (NISQ) era~\cite{Preskill2018}. From that perspective, one can benefit from 
a microscopic modelling of these devices to partially alleviate the stringent resource requirements of GST. In this work, instead of using a device-agnostic approach, we advocate for a microscopically-motivated parametrization of the  gate set~\eqref{eq:gate_set},  having a specific setup in mind: gates mainly limited by coloured phase noise. As discussed in~\cite{velazquez2024dynamical}, the effect of frequency and phase stochastic fluctuations of current trapped-ion QIPs can be accurately described with the filter-function formalism~\cite{Kofman2000,PhysRevLett.87.270405,PhysRevLett.93.130406,Gordon_2007,Uhrig_2008,PhysRevB.77.174509,Biercuk_2011,Almog_2011}. We use this formalism to demonstrate that the  imperfections of the gate set can be related to various filtered integrals of the noise spectral density, reducing considerably the number of parameters that is required in GST. In particular, for the gate set under study in this work, the total number of parameters for fully-general approach amounts to $N_{GST}=67$. This  contrasts with that 
required to characterize the gate set under the coloured phase noise, where  $N_{GST}=11$ is required for a Markovian estimation or $N_{GST}=16$ for a non-Markovian one. We do also show that it is straightforward to account for amplitude noise fluctuations as well, at the cost of just two additional parameters in order to completely characterize the gate set. Finally, we refer to the GitHub repository \texttt{ColouredGST} \cite{myrepo} as a guide for the numerical implementation of the results in this work.

This manuscript is organized as follows: Sec. \ref{sec_1} presents the theoretical description of our parametrized GST. 
We delve into some peculiarities derived from this microscopically-motivated approach, such as the simplification of the circuit selection process and the absence of the so-called gauge redundancy, which is a characteristic of fully-general approach to GST~\cite{Nielsen_2021, 
greenbaum2015introduction,
blumekohout2013robust,
PhysRevA.89.052109}. Sec. \ref{sec_2} constitutes a numerical validation of our parametrized GST. By considering specific numerical simulations of the coloured phase noise, we can compare the GST estimates $\hat{\mathcal{G}}$ to the exact microscopic operations $\bar{\mathcal{G}}$ that stem from the stochastic averages over the phase fluctuations. In this way, we can analyse the accuracy of our estimation and discuss its performance for different noise regimes, including the effect of time correlations and non Markovian evolution as well as the amplitude noise extension of the model. Moreover, we compare the results of our parametrized  GST approach with those obtained from fully general GST applied to the same numerically-simulated data. We close this manuscript in Sec.~\ref{conclusions} with our conclusions, also mentioning some future research lines motivated by this work.

\section{\bf Gate set tomography for coloured phase noise}
\label{sec_1}
The goal of this section is to give a comprehensive account of our customized GST, comparing the parameterization for the phase-noise model on single-qubit gates to   the standard protocol of long-sequence GST. For a  review of important aspects of GST, see Appendix~\ref{appendix_2}, where we focus on  the concepts used throughout this work, which are  discussed in more depth  in the review by Nielsen et {\it al.} \cite{Nielsen_2021} for  fully-general GST. As outlined in the introduction, the main advantage of employing a microscopically-motivated model lies in the reduction of the parameters required for a complete self-consistent characterization. 

Although approached from a  different angle,  using a simplified model to efficiently learn the noise of a device is successfully exemplified by \cite{van_den_Berg_2023}, where one  restricts the estimation to  a Lindblad-Pauli channel~\cite{10.1145/3408039,fawzi2023lower}. In fact, for a Lindbladian evolution, one can estimate the generators of the channel rather than the channel itself, which allows to reconstruct the full dynamical quantum map~\cite{Childs2001,PhysRevA.67.042322,Howard_2006,Bairey_2020,PRXQuantum.3.030324,frança2022efficient,PhysRevApplied.17.054018,PhysRevApplied.18.064056,PhysRevA.101.062305}, and obtain advantages from compressed sensing when the microscopic noise generators have a clear structure with leading components~\cite{dobrynin2024compressedsensing}. In some situations, however, time correlations in the noise and non-Markovian effects forbid this approach. Moreover, the Pauli-type channel may also be  a crude approximation to the actual native gate set. As shown in this work, even in this situations, one can still benefit from a detailed microscopic modelling of the non-Lindblad and non-Pauli channel through a reduced  parametrization of GST.  This reduction not only leads to significant savings in computational resources, but also results in a proportional decrease in the number of experiments required. Moreover, as we describe now in detail, our approach circumvents certain  challenges encountered in a fully-general GST. 

\subsection{Filtered-noise parametrization of the gate set}
\label{mic-parameterization}

As discussed in the introduction and in Appendix~\ref{appendix_2}, GST considers a noisy gate set~\eqref{eq:gate_set}, which can be described using the {process matrix} representation for the quantum channel associated to each gate 
\begin{equation}
\label{eq:chi_matrix}
\mathcal{E}_i(\rho) = \sum_{\alpha, \beta}\chi^{(i)}_{\alpha,\beta}E^{\phantom{\dagger}}_{\alpha}\rho E_{\beta}^\dagger,\hspace{2ex}i\in\{1,\cdots,N_G\}.
\end{equation}
Here, $\lbrace E_{\alpha}: \alpha\in\{1,\cdots,d^2\}\rbrace$ can be any basis of the space of linear operators $\mathsf{L}(\mathcal{H})$, such as the aforementioned Pauli basis~\cite{nielsen_chuang_2010}. The above set of $\{\chi^{(i)}\}$ matrices must be semidefinite positive and fulfill a so-called trace constraint~\eqref{eq:trace_constraint}. 
In full generality, each of them must be parametrized via $d^2(d^2-1)$ real parameters. The fiducial operators $\rho_0,M_0\in\mathcal{G}$ require $N_{\rm SPAM}=d^3-1$ real parameters, leading to a total of $N_{GST}=N_G\times d^2(d^2-1)+d^3-1$ parameters to be estimated in GST. 

Rather than  treating the QIP as a black box and aiming at GST with full generality, we advocate for a specific  parametrization  that employs a physically-motivated microscopic modelling of the gates. 
To define a gate set, 
let us start by considering the case of perfect unitary gates 
\beq
\label{eq:ideal_gate}
G^{\rm id}_i(\theta_i,\phi_i)={\rm exp}\left\{-\ii\frac{\theta_i}{2}\big(\cos\phi_i\sigma_x-\sin\phi_i\sigma_y\big)\right\},
\eeq
where $\theta_i (\phi_i)$ is the pulse area (phase). In this work, we will slightly abuse the notation by using the same symbol $G_i$ for the unitary gate, its Pauli transfer matrix representation, and the gate label.  IC can be achieved  by using a single $\theta=\pi$ pulse with $\phi=0$  that maps $\ket{0}\!\bra{0}\mapsto\ket{1}\!\bra{1}$, and a pair of $\theta=\pi/2$ pulses with $\phi=0$ and $\phi=3\pi/2$, which map $\ket{0}\!\bra{0}\mapsto\ket{+}\!\bra{+}$ and $\ket{0}\!\bra{0}\mapsto\ket{+{\rm i}}\!\bra{+{\rm i}}$, respectively, where $\ket{\pm w}=(\ket{0}\pm w\ket{1})/\sqrt{2}$. These $\pi/2$ pulses also suffice to rotate the measurement basis from $b=z$ to $b=y,x$, achieving the required IC. In the context of  long-sequence GST, it will be convenient to enlarge this gate set to $N_{G}=5$ by also including the inverse of the $\pi/2$-pulses, as this does not require additional parameters when considering the microscopic noisy gates, and can lead to a higher sensitivity of the circuits to the parameters that long-sequence GST aims at estimating. We thus define our gate set 
\begin{equation}
\label{eq: gate_set}
\begin{split}
\mathcal{G}&=\Bigl\{\rho_{0},\hspace{1ex}\big\{ G_i(\theta_i,\phi_i), \,i={1,\cdots,5}\big\},\hspace{1ex}M_0\Bigr\}, 
\end{split}
\end{equation}
where we have introduced the following angles
\beq
\label{eq:gate_ste_angles}
(\theta_i,\phi_i)\in\left\{(\pi,0),\left(\frac{\pi}{2},0\right),\left(\frac{\pi}{2},\frac{3\pi}{2}\right),\left(\frac{\pi}{2},\frac{\pi}{2}\right),\left(\frac{\pi}{2},\pi\right)\right\}.
\eeq

In an experiment, the gates $G_i$ will deviate from the ideal unitaries~\eqref{eq:ideal_gate}. One common source of noise that appears in several platforms and limits the gate fidelities is that of phase noise, which can be modelled by a particular stochastic process and characetrised by its experimentally-measurable power spectral density $S(\omega)$. As detailed in Appendix~\ref{appendix_1}, using a dressed-state master equation for the description of the noisy gates, one can find a parametrization of  the process matrices~\eqref{eq:chi_matrix} that can be  specified by only four real parameters $\chi^{(i)}\!\left(\Gamma_{1}(t_i),\Gamma_{2}(t_i), \Delta_{1}(t_{i}),\Delta_{2}(t_{i})\right)$. In turn, these parameters incorporate information of the noise and the single-qubit driving leading to the gate, which are encapsulated in    filtered integrals of the power spectral density (PSD) 
\beq
\label{eq:Gammas}
\begin{split}
	\Gamma_n(t_i) = & \int_{-\infty}^\infty \!\!{\diff\omega}\,S(\omega)\,F_{\Gamma_n}(\omega,\Omega_i,t_i),\\
	\Delta_n(t_i) = & \int_{-\infty}^\infty \!\!{\diff\omega}\,S(\omega)\,F_{\Delta_n}(\omega,\Omega_i,t_i).
\end{split}
\eeq
Here, the  filter functions $F_{\Gamma_n}(\omega,\Omega_i,t_i), F_{\Delta_n}(\omega,\Omega_i,t_i)$ are explicitly given in Eqs.~\eqref{eq:Gamma1_filter}-\eqref{eq:noN_{t}arkov_filter_functions}, where we note that $t_i (\Omega_i)$ stands for the pulse duration (Rabi frequency) of each gate. It is interesting to  observe that the noisy gates $G^{\rm id}_i(\theta_i,\phi_i)\mapsto G_i(\theta_i,\phi_i)$ in Eq.~\eqref{eq: gate_set} with shared pulse duration $t_{i}$ and  Rabi frequency $\Omega_i$, such that the pulse area $\theta_i=\Omega_i \, t_{i}$ is the same, are actually described  by the same filtered noise parameters under our noise modelling and, furthermore, these parameters do not depend on the driving phases $\phi_i$ in Eq.~\eqref{eq:gate_ste_angles}. The only change for different gates is that the microscopic parameters enter in different places of the corresponding process matrices, as detailed in Eqs.~\eqref{eq:chi_app}-\eqref{eq:chi_app_3pi_2}. Hence, no additional effort will be necessary to characterize $G_{3}(\pi/2,3\pi/2)$, $G_{4}(\pi/2,\pi/2)$, $G_{5}(\pi/2,\pi)$ via GST under the assumption that $G_1(\pi,0),G_2(\pi/2,0)$ can be estimated efficiently. In fact, 
including these operations in the gate set can lead to an improved parameter sensitivity in long-sequence GST.

Let us  note that the standard approach of GST emphasises the requirement of a Markovian evolution during  the noisy  gates, and a failure in the gate set reconstruction can sometimes be associated to a model violation due to the breakdown of the  the Markovianity assumption~\cite{Nielsen_2021}. On the other hand, when considering the full PSD of the phase noise~\eqref{eq:Gammas}, we are indeed incorporating  finite time correlations~\eqref{eq:autocorrrelation_function}, and thus deviating from a purely Markovian prediction base on a Lindbladian approach. Hence, the discussion about Markovianity and GST is slightly more succinct. As detailed in~\cite{velazquez2024dynamical}, one can rigorously connect the values of  the filtered noise parameters $\{\Gamma_n(t_i), \Delta_n(t_{i}) \}$ to a   non-Markovianity measure~\cite{Rivas_2014,RevModPhys.88.021002,RevModPhys.89.015001,LI20181,CHRUSCINSKI20221} that is based on the lack of a completely-positive division of the resulting quantum dynamical map~\cite{PhysRevLett.105.050403,PhysRevA.89.042120}. If we include $\Gamma_2(t_i), \Delta_2(t_{i})$ in the noisy gate set parametrization, and the GST estimates that these parameters are indeed non-zero, the reconstructed gate set corresponds strictly to a non-Markovian quantum evolution. Hence, GST can indeed estimate non-Markovian gates, albeit assuming that there are no time correlations in between the SPAM operations and the different gates in the GST sequence, i.e. we  assume that the noisy parametrization of   each gate is independent of the previous history of gates applied within a given sequence. Considering that the coloured noise will have a characteristic   correlation time $\tau_{\rm c}$, we are thus assuming that $t_i/\tau_{\rm c}$ can be sufficiently large so that non-Markovian effects take place during each gate, but the time in between consecutive gates $\Delta t_{i,j}\gg\tau_{\rm c}$, such that one can neglect correlations between different gates.

In principle, one could exploit the formalism of the noise filtered integrals to extend the modelling of fluctuations also in between gates, accounting for non-Markovianity in the full gate sequence. However,  the resource cost of GST would increase considerably, as each circuit of the long-sequence GST would then depend on the specific gate history, and thus require even more parameters to be estimated. Some proposals of non-Markovian tomography have already been made employing tensor networks \cite{PRXQuantum.3.020344, li2023nonmarkovian}, always at the expense of an even greater computational cost. We leave a detailed study of these questions  for future work.

In most of the discussion below, we will show that  the leading-order effect of the noisy gate set  can be characterized by restricting to a non-zero $\Gamma_1(t_i)$ and $\Delta_1(t_i)$, whereas $\Gamma_2(t)=\Delta_2(t)=0$. It is worth noting that, under this Markovian assumption,  the total number of parameters -- and minimal number of experiments -- required to completely characterize this gate set decreases from $N_{GST} = 67$ when working with the fully general description of the channels to $N_{GST} = 11$ with our parametrized approach.
{In  the last part of Sec.~\ref{num_results}, we present a generalization of our parametrized GST to allow for non-zero  $\{\Gamma_2(t_i), \Delta_2(t_i)\}$, where we quantify how certain noise regimes require GST to account for these non-Markovian effects in order to achieve higher accuracies.}

\subsection{Microscopic parameters and  gauge redundancy}
\label{sec: gauge_redundancy}

Once the gate set and its parametrization have been described, we can  discuss the different estimation strategies for GST. As reviewed in Appendix~\ref{appendix_2}, the first schemes of GST~\cite{blumekohout2013robust} considered a linear-inversion scheme~\eqref{eq:linear_GST} that only requires measuring the probabilities of simple circuits composed of the SPAM operations and the  action of the individual gates  $G_i\in\mathcal{G}$ (see Fig.~\ref{fig: tomography_scheme}\textcolor{magenta}{c}). This approach leads to a certain redundancy in the estimated gate set $\hat{\mathcal{G}}$ found by linear GST. As discussed in that Appendix, it is easy to see in the super-operator formalism in which states and measurements are vectorized $|\hat{\rho}_0\rangle\!\rangle, \langle\!\langle \hat{M}_{0}| $, and quantum channels have a matrix representation $\hat{G}_i$, that one can apply a similarity transformation $T$ to obtain a new gate set $\hat{\mathcal{G}}'$ that is equally valid when one transforms simultaneously  the estimated noisy gates and the  fiducial state and measurement operators 
\begin{equation}
\begin{split}
|\hat{\rho}_0\rangle\!\rangle \rightarrow T|\hat{\rho}_0\rangle\!\rangle,\hspace{1ex}
\hat{G}_i\rightarrow T\hat{G}_iT^{-1}, \hspace{1ex}
\langle\!\langle \hat{M}_{0}|\rightarrow \langle\!\langle \hat{M}_0|T^{-1}.
\end{split}
\end{equation}
 This redundancy in linear GST is referred to as gauge freedom, and  arises due to the absence of a privileged reference frame, which is the price to pay when transitioning from QPT to GST. Moreover, it is actually not exclusive to GST but  also appears in other  characterization techniques such as randomised benchmarking~\cite{Proctor_2017}.

When using a parametrized gate model, this redundancy  may be partially or even completely eliminated, as the microscopic modelling can actually define a privileged reference frame. This is precisely the situation for our Markovian parametrization, which only depends on the filtered-noise parameters 
that do not change under any similarity transformation. Under this transformation, the matrix structure of our parametrization of the noisy gates must be preserved, as it describes the whole family of imperfect gates subject to the coloured phase noise, albeit with a different set of parameters $\{\Gamma_1(t_i),\Delta_1(t_1)\}\mapsto\{\Gamma'_1(t_i),\Delta'_1(t_1)\}$. Considering that  under a similarity transformation $\hat{G}_i'=T^{-1}\hat{G}_iT$, both $\hat{G}_i$ and $\hat{G}_i'$ must possess the same spectrum, one can readily see that our parametrization does not allow for gauge redundancy.
By calculating the eigenvalues of the corresponding noisy gates, we find that $\sigma(G_{i}(\Gamma_1, \Delta_1))\neq \sigma(G_{i}(\Gamma_1', \Delta_1'))$ for any physically-admissible $(\Gamma_1',\Delta_1')$, where 
$\sigma(O)$ symbolizes the spectrum of the operator $O$. This indicates that there is only one pair of parameters consistent with the GST probabilities.

The absence of gauge redundancy renders unnecessary the need for gauge optimization techniques used in fully-general GST~\cite{Nielsen_2021}. Typically, these techniques are required to compute distance metrics, which depend on the  gauge choice, adding further complexity to the already demanding estimation process of GST. Typically, the gauge-fixed gate set is defined as the one that is closest in a certain notion distance to the ideal unitary gate set~\eqref{eq:ideal_gate}, while being compatible with the measured frequencies. Due to its simplicity, we will opt for trace distance $T(\rho,\rho')=\half\sqrt{(\rho-\rho')^{\dagger}(\rho-\rho')}$  to get a measure of `closeness' between a reconstructed gate set element and a target one~\cite{nielsen_chuang_2010}. Note that the (in-)fidelity~\cite{Jozsa}  is not a wise choice  in this context, as it is only well-behaved for normalized density matrices, and not every element in our gate set satisfies this condition. For our parametrization, the trace distance between two gates is   
\begin{equation}
\label{eq:td}
\begin{split}
T_{\hat{G},\bar{G}}\!=\abs{\frac{\ee^{-\hat{\Gamma}_{\!1}}-\ee^{-\bar{\Gamma}_{\!1}}}{4}\!}
+\half\!\biggl(\!\!\ee^{-\hat{\Gamma}_{\!1}}+\ee^{-\bar{\Gamma}_{\!1}}-\biggr.\left.2\ee^{-\frac{\hat{\Gamma}_{\!1}+\bar{\Gamma}_{\!1}}{2}}\!\!\cos\!\left(\!\frac{\hat{\Delta}_{\!1}-\bar{\Delta}_{\!1}}{2}\!\right)\!\!\!\right)^{\!\!\!\!\frac{1}{2}}\!\!\!,
\end{split}
\end{equation}
which will be useful to estimate the error in our parametrised GST, where we will compute an average trace distance for all elements in the gate set (trace distances for the fiducial elements are introduced at the end of the next subsection). 
We do not need to delve here into whether gauge fixing by minimizing the trace distance to the unitary gates  is a fully-consistent technique or not, as one could imagine sufficiently-noisy instances in which, by minimizing the distance to the unitary gate set  along a gauge orbit, one is actually moving further apart from the actual noisy channels that represent the imperfect gate set. This problem is entirely absent in our case, as there is no gauge redundancy left in our parametrization.

\subsection{ Long-sequence maximum-likelihood estimation}

Let us now discuss the posterior GST developments that superseded  linear GST \cite{blumekohout2013robust},  showing that one can improve the accuracy and precision of linear GST estimates by allowing for  sequences of gates forming more complex circuits. This leads to the concept of long-sequence GST (see Fig.~\ref{fig: circuits}) briefly discussed in Appendix~\ref{appendix_1}.   Let us introduce $\gamma_{s}$, $\gamma_{c}$, and $\gamma_{b}$ as labels of the different state preparation, base circuit and measurement basis components of a particular circuit $\boldsymbol{\gamma}=(\gamma_s,\gamma_c,\gamma_b)$, respectively. The probability characterizing the measurement outcomes of each of the $N_{C}$   circuits in long-sequence GST can be expressed in terms of the corresponding product of parametrized  process matrices 
\begin{equation}
\label{eq: probabilities}
p_{\boldsymbol{\gamma}}(m_b) = \sum_{\boldsymbol{\omega}}\,
\chi^{\gamma_{b}}_{\alpha,\beta}\chi^{\gamma_{c}}_{\tau,\zeta}
\chi^{\gamma_{s}}_{\epsilon,\delta}\mathrm{Tr}\left\{M_{0,m_b}E_{\alpha}^{\phantom{\dagger}}E_{\tau}^{\phantom{\dagger}}E_{\epsilon}^{\phantom{\dagger}}\rho_{0}E_{\delta}^{\dagger}E_{\zeta}^{\dagger}E_{\beta}^{\dagger}\right\},
\end{equation}
where $\boldsymbol{\omega}=(\alpha,\beta,\tau,\zeta,\epsilon,\delta)\in\{1,\cdots,d^2\}$ is a super-index that labels the components of the process matrices in the Pauli basis for each part of the GST circuit.  We recall that these matrices are expressed as products of the parametrized $\chi^{(i)}$ matrices of the individual noisy gates
in Eqs.~\eqref{eq:chi_app}-\eqref{eq:chi_app_3pi_2}, each of which has  a non-linear dependence on the noise-filtered integrals~\eqref{eq:Gammas}  $\chi^{(i)}\!\left(\{\Gamma_n(t_i), \Delta_n(t_{i}) \}_{n=1,2}\right)$. We group all of these parameters, together with those of the fiducial operations, in a single vector $\boldsymbol{\theta}$, which will allow us to formalise the GST as a particular problem of multi-parameter point estimation~\cite{rossi2018mathematical}.

\begin{figure}
  \centering
  \includegraphics[width=1\columnwidth]{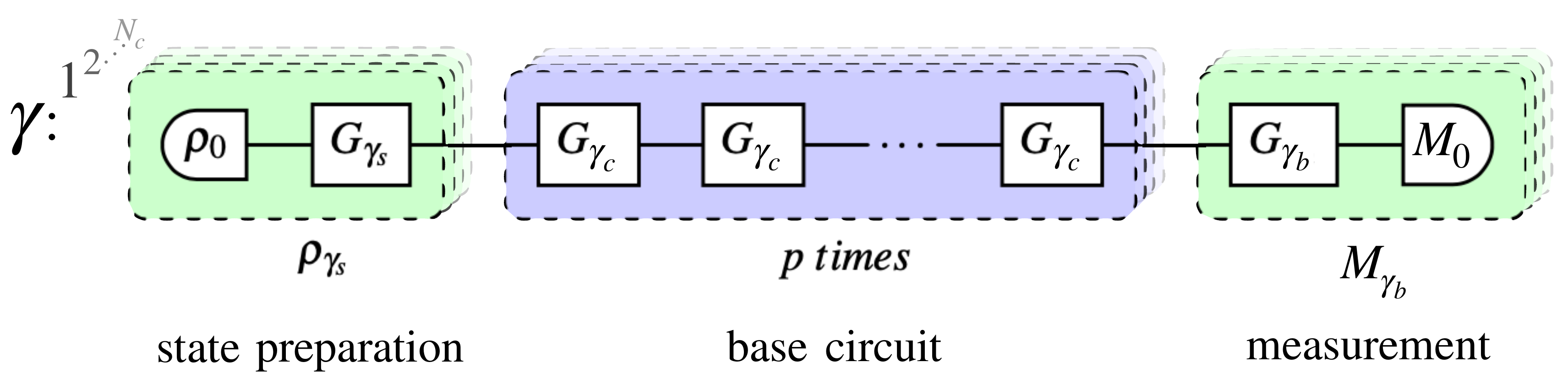}
  \caption{\textbf{Circuit design for long-sequence GST:} The conventional long-sequence GST scheme uses $N_c$ different possible combinations of circuits, here labeled by $\boldsymbol{\gamma}$, that contains the state preparation $\gamma_s$, base circuits $\gamma_c$, and measurement $\gamma_b$ sublabels. When preparing IC sets of initial states $\rho_{\gamma_s}$ and POVM elements $M_{\gamma_b}$, we rely on the action of single gates. For our parametrization, the base circuit germs germs $G_{\gamma_c}$ are also single gates, as these are a valid amplificationally complete set (see Sec \ref{circuit_selection}). Each circuit results in $d$ possible outcomes resulting from the binary Z-basis measurement for each qubit.}
  \label{fig: circuits}
\end{figure}

To avoid non-physical estimations in GST, one defines a cost function that depends on the differences between these parametrized circuit probabilities and the corresponding measured relative frequencies $f_{\boldsymbol{\gamma}}(m_b)$, which is then minimised under specific constraints  to restrict to  physically-admissible gate sets. Following some of the ideas used in  fully-general GST~\cite{Nielsen_2021}, we integrate  a negative log-likelihood cost function $\mathcal{C}_{\rm ML}(\boldsymbol{\theta})$ with a weighted least-squares cost function $\mathcal{C}_{\rm LS}(\boldsymbol{\theta})$ throughout the iterative process that minimizes  
\beq
\label{MLE_GST}
\begin{split}
&\hat{\boldsymbol{\theta}}=\texttt{argmin}\hspace{1ex} \begin{cases}
                    \mathcal{C}_{\rm ML}(\boldsymbol{\theta})=-\sum\limits_{\boldsymbol{\gamma}}\sum\limits_{m_b}f_{\boldsymbol{\gamma}}(m_b)\log p_{\boldsymbol{\gamma}}(m_b),\\
                    \mathcal{C}_{\rm LS}(\boldsymbol{\theta})=\sum\limits_{\boldsymbol{\gamma}}\sum\limits_{m_b}\frac{1}{p_{\boldsymbol{\gamma}}(m_b)}\big(f_{\boldsymbol{\gamma}}(m_b)-p_{\boldsymbol{\gamma}}(m_b)\big)^2\!.
                    \end{cases}\\ 
\end{split}
\eeq
In our optimization routine, we choose circuits with a maximum depth that is logarithmically spaced, and run an optimization iteration for each of them, incorporating data from lower depths into the larger-depth subsequent steps. 
Additionally, the solution obtained from each iteration serves as the initial guess for the subsequent one, using the ideal unitary gate set  as the initial guess for the first iteration. This workflow, originally developed for fully-general GST,  prevents us primarily from getting stuck at an undesirable local minima of the cost function, but also from increasing the potential `branch ambiguity' with depth. This problem occurs when any of the parameters under estimation is multi-valued, such that $\theta\rightarrow\theta\hspace{1ex}\rm{mod}$$K$. Then, the Heisenberg-like increase of accuracy stemming from considering deeper versions of the circuits also increases the multi-value ambiguity, so that one has $\theta/p\rightarrow\theta/p\hspace{1ex}\rm{mod}$$K/p$. Luckily, the inclusion of the previously mentioned logarithmically spaced maximum depth circuits is sufficient to discern the `true' branch. Regarding the cost function, we opt for $\mathcal{C}_{\rm LS}$ during the optimization involving  the initial low-depth stages, while $\mathcal{C}_{\rm ML}$ is employed at later stages when a sufficiently accurate estimation  has already been reached at  previous iterations that can be used for the initial guess of the subsequent ones, as this can correct for the potential biased behaviour of $\mathcal{C}_{\rm LS}$ for rare events \cite{Nielsen_2021}. We remark that the choice among the cost functions is rather heuristic, and may vary depending on the structure of the problem. 

So far, the definition of the optimization in Eq.~\eqref{MLE_GST} is not complete, as we have not detailed yet the constraints. Working with the process matrix representation~\eqref{eq:chi_matrix}, we need to impose the positive semi-definite and trace preservation constraints in Eq.~\eqref{eq:trace_constraint}. These conditions manifest neatly with our approach, as our parametrized process matrices are always trace preserving, while the only condition required to ensure complete positivity in the Markovian regime reduces to $\Gamma_1(t_{i})\geq 0$. Additionally, we restrict $\rho_0$ and $M_{0}$ to be positive semi-definite, and set the trace of the fiducial state to be unity. Consequently, the  minimization~\eqref{MLE_GST}  must be constrained as
\beq
\begin{split}
\texttt{subject to} \hspace{1ex} \begin{cases} \Gamma_1(t_i)\geq 0, \hspace{4ex} i=1,...,N_{G},\\
\sigma ( \rho_0 ) \in\mathbb{R} ^+,\hspace{2ex} \mathrm{Tr} \lbrace \rho_0 \rbrace = 1, \\
\sigma( M_{0}) \in\mathbb{R}^+.
\end{cases}
\end{split}
\eeq
From this point onward, we adopt the orthonormal Pauli basis
with $N=1$ for single-qubit systems. With this basis, the native state and measurements can be expressed as $\rho_0 = \frac{1}{\sqrt{2}}\left(E_0+\sum_{{\alpha}=1}^{3}r_{\alpha}E_{\alpha}\right)$ and $\hspace{2ex}M_{0}=\frac{1}{\sqrt{2}}\left(\sum_{{\alpha}=0}^{3} e_{\alpha}E_{\alpha}\right)$ respectively, where $r_{\alpha}$, $e_{\alpha}$
are real coefficients. Then, constraints on $\rho_0$ and $M_0$ can be further simplified to 
\begin{equation}
\begin{split}
\sum\limits_{\alpha=1}^{3}r_{\alpha}^2&\leq 1,\\
e_{0} \in [0,2],\hspace{2ex}\sum\limits_{\alpha=1}^{3}(e_{\alpha})^2&\leq \min\lbrace (e_{0})^2, (2-e_{0})^2\rbrace.
\end{split}
\end{equation}

In contrast to fully-general GST~\cite{Nielsen_2021}, in which the CPTP conditions are not imposed when working with the Pauli transfer matrix representation of a quantum channel (see Appendix~\ref{appendix_1}), the former physical constraints are notably simpler and can be readily imposed during the optimization stage independently of the channel representation. {Note that, in parallel with the expression in Eq.~(\ref{eq:td}), these parametrizations for the fiducial elements lead to the simple trace distances $T_{\hat{\rho}_0,\bar{\rho}_0}=\sqrt{\sum_{\alpha=1}^3(\hat{r}_{\alpha}-\bar{r}_{\alpha})^2}$ and $T_{\hat{M}_0,\bar{M}_0}=\sqrt{\sum_{\alpha=1}^3(\hat{e}_{\alpha}-\bar{e}_{\alpha})^2}$.} 

\subsection{Selection of parametrized base circuits }
\label{circuit_selection}
Having covered the main aspects of our parametrized approach to GST, we can now discuss how to  select a set of base circuits that permit achieving an increased sensitivity in the estimation of  the filtered noise parameters. The efficacy of long-sequence GST hinges on the capability of amplifying each model parameter in $\boldsymbol{\theta}$ by the repetition of base circuits $p\gg 1$ times. This enables a reduction of the estimation uncertainty  $\Delta\theta_p = {\Delta\theta}/{p}$, where $\Delta\theta$ denotes the uncertainty for a single repetition $p=1$. This  contrasts  the reduction of the shot-noise uncertainty mentioned in the introduction, which scales with  the number of shots  used to infer the relative frequencies as $1/\sqrt{N_{\rm shots}}$. This idea, first explored in \cite{blumekohout2013robust},  has  evolved into its current form  described in \cite{Nielsen_2021} for  fully-general GST, including  steps  of so-called circuit/germ selection and  fiducial pair reduction that we now consider for our parametrized GST.

In long-sequence GST, one has to carefully determine a set of {germs} -- circuits composed of gates from the gate set -- that suffice to amplify each gate-set parameter    when repeated $p$ times. These sets,  coined {amplificationally complete} (AC), rarely result in a one-to-one correspondence between circuits and parameters, as there are  many circuits with specific combinations of gates that end up amplifying common gate-set parameters. In addition, evaluating the outcome of every of these germ circuits for  all  SPAM operations on the fiducial pair of initial state and measurement $\rho_0,M_0$ increases the total number of circuits considerably. This can actually waste resources in terms of experimental  shots for certain measurements that do not give more information about the parameters. In essence,  there are redundant combinations of base and SPAM circuits that increase the number of experiments beyond a minimal GST design, and it  is thus  important to develop techniques for circuit selection (CS) and fiducial pair reduction (FPR) to  reduce this redundancies~\cite{Nielsen_2021}. In this way, the  measurement resources can be optimally used for those AC circuits that provide more information gain.

Interestingly, we can actually avoid this two-step CS and FPR process by demonstrating that our parametrized gates already constitute an AC set, and there is no need to combine them in more complicated circuits. This enables us to effortlessly identify minimal GST designs that are  optimal for single-qubit gates. Finally, as explained before, our restricted model does not suffer from gauge redundancy,  obviating the need for additional gauge optimization techniques to establish a common reference frame to compute distance metrics.

In fully-general GST, one identifies an AC set of germs $g_i= G_{i_1}\circ G_{i_2}\circ...\circ G_{i_N}$ by finding non-vanishing quantities for 
\begin{equation}
\label{eq:der_germs}
\boldsymbol{\nabla}_{i}^{(\infty)}=\lim_{p\to\infty}\frac{1}{p}{\boldsymbol{\nabla}_{\boldsymbol{\theta}}{\tau^p(g_i)}}, \hspace{3ex}i=1,...,N_{germs}
\end{equation}
where $\tau(g_i)$ stands for the  Pauli transfer matrix  for the germ $g_i$. 
The general approach to calculate this amplification gradient  uses Schur's lemma to establish a relationship between the expression above and the derivative when $p=1$, under the assumption that $\tau(g_i)$ is approximately unitary \cite{Nielsen_2021, ostrove2023nearminimal}. This assumption remains  valid for the high-fidelity QIPs until the number of repetitions of the germs is too large, and the  accumulated error starts to dominate leading to a deviation from the unitary behaviour. This transition typically occurs at depths that are  considerably larger than ${1}/{\epsilon}$, where $\epsilon$ is defined as the {decoherence rate per gate}. In combination with experimental limitations, this error accumulation prevents one from  using  arbitrarily long circuits in GST to enhance the estimation accuracy and precision.

Fortunately, due to the simplicity of our parametrized expressions for the noisy gates, we can evaluate the expression~\eqref{eq:der_germs} analytically. Furthermore, considering single-gate germs with a Markovian parametrization of the phase noise, we find $[G_i(\Gamma_1(t_i), \Delta_1(t_i))]^p = G_i(p\Gamma_1(t_i), p\Delta_1(t_i))$, which is indeed the amplification behaviour one looks for when executing long sequence GST. {Coming back to Eq.~\eqref{eq:der_germs}, we note that each of the non-vanishing matrix elements in this limit, denoted here by the indices $(j,k)$, satisfies 
\begin{equation}
\label{eq:der_germs_P-N}
\big(\boldsymbol{\nabla}_{i}^{(\infty)}\big)_{j,k}\approx \ee^{-p\Gamma_1(t_i)},
\end{equation}
}
leading directly to two conclusions. Firstly, as anticipated by the linear amplification of parameters with $p$, the individual gates indeed serve already as valid germs for long-sequence GST. This becomes evident by evaluating Eq.~\eqref{eq:der_germs_P-N} in the high-fidelity regime ($p\Gamma_1(t_i)\ll 1$). Typically, short germs are prioritized to maximize precision when constrained by a maximum circuit depth, so this is significant. In addition, as we are not combining different gates to form germs, the sets of parameters that different germs amplify are disjoint. Secondly, as we will discuss latter, we can use this limit to predict the region in which the error dominates, corresponding to {$p\gtrsim{1}/{\Gamma_1(t_i)}$}, where long-sequence GST will not lead to further improvements on the accuracy and precision of the estimation. It is significant to mention that individual gates are no longer sufficient to perform long sequence GST when the Markovian approximation is relaxed ($\Gamma_2(t_i),\Delta_2(t_i)\neq0$), as those parameters need more complex combinations of gates to exhibit amplification with depth. Nevertheless, as mentioned at the end of Subsec.~\ref{mic-parameterization}, we recall that there are situations, as those explored in this paper, for which the contributions introduced by these parameters can be neglected. 

Once germs have been identified, one only needs to incorporate fiducial pairs to determine the complete circuits that will be used in our routine. Traditionally, in fully-general GST,  all germs would be used -- which might probably be redundant -- and only a reduced number of fiducial pairs would be applied to mitigate further redundancies. Alternatively, {per-germ fiducial pair reduction (FPR)} \cite{ostrove2023nearminimal} is a process that selects germs to minimize redundancies beforehand,  resulting in near-minimal GST designs. However, it requires solving an NP-complete problem, a two-step optimization involving {column subset selection problem}~\cite{shitov2017column}, which becomes already challenging for moderately large values of $N_p$.
Our parametrization presents a structure similar to that found in per-germ FPR, but avoids the column subset selection problem. First, we associate each gate $G_i$ with the reduced set of parameters it amplifies which, as we anticipated, forms a  set that is disjoint with respect to all the other gate-amplified parameter sets. Subsequently, we perform a small variation from the unitary values of each of these parameters. Then, for each of them, we select the fiducial pair that is more sensitive to the specific variation given the base circuit built from $G_i^p$. We are justified in the computation of this at each $p$ due to the simple analytic form of $G_i^p$. This process can be repeated for every parameter to find a one-to-one correspondence between parameters and circuits. In other words, we find that $N_p$ circuits suffice to determine and amplify each of the $N_p$ gate-set parameters from $\boldsymbol{\theta}$, constituting a minimal GST design.


\section{\bf Numerical benchmarks with stochastic  noise}
\label{sec_2}
After discussing all the central aspects of our parametrized GST, we now  present a numerical benchmark, testing it with simulations and comparing its performance against fully-general GST. To run our parametrized GST, we rely on the practical open-source python software \texttt{pyGSTi} \cite{Nielsen_2020, pygsti}, using stochastic numerical simulations to obtain the noisy data. When adapting GST to our parametrized formulation, we integrate a low-level implementation of GST with our model's specific requirements, all of which were detailed in the previous section. We have independently programmed the GST maximum-likelihood estimation from the start, which has served to test that the low-level integration of the parametrization of \texttt{pyGSTi} is working seamlessly. Ultimately, we rely on this open-source approach, as it offers extensive and useful capabilities. This is specially convenient when comparing  with the fully-general GST, as the package is designed to offer high-level functionalities to work easily within this `agnostic' framework. When adapting GST to our microscopic noise model, the benefits are partially reduced in comparison to our original GST implementation due to the lower-level implementation required, although these are still appreciable. The \texttt{pyGSTi} implementation of the model in this work, used to obtain the main results of it, can be accessed through the GitHub repository \texttt{ColouredGST} \cite{myrepo}.

\subsection{Parametrized GST for Ornstein-Uhlenbeck noise}
\label{num_results}
To validate the accuracy of our results, we simulate the dynamics of the set of quantum channels~(\ref{eq: gate_set}) under {Ornstein-Uhlenbeck}  phase noise \cite{gardiner1985handbook, 10.1119/1.18210}. This choice is motivated by the facility with which we can tune the correlation time $\tau_{\rm c}$ and the diffusion constant $c$ of this process, which fully define the Langevin  stochastic differential equation 
\beq
\label{eq:OU_langevin}
\frac{{\rm d}}{{\rm d}t}\tilde{\delta}(t)=-\frac{1}{\tau_{\rm c}}\tilde{\delta}(t)+\sqrt{c}\tilde{\xi}(t),
\eeq
Here, the random process $\tilde{\delta}(t)$ models a coloured phase noise of the driving~\eqref{eq:stoch_H_Appendix} leading to single-qubit gates, and $\tilde{\xi}(t)$  stands for a white noise seed of the fluctuations that ultimately  leads to differences with respect to the ideal unitaries~\eqref{eq:ideal_gate}. A useful feature of the Ornstein-Uhlenbeck process is that there is a closed analytical expression for its stochastic trajectories which,  discretizing time as $t_n=t_0+n\Delta t$, reads as follows
\begin{equation}
\label{eq:traj_OU}
    \tilde{\delta}(t_{n+1})=\tilde{\delta}(t_n)\ee^{-\frac{\Delta t}{\tau_{\rm c}}}+\sqrt{\frac{c\tau_{\rm c}}{2}\left(1-\ee^{-\frac{2 \Delta t}{\tau_{\rm c}}}\right)}\tilde{u}_{n}.
\end{equation}
Here, we have introduced  a unit normal random variable per time step $\tilde{u}_n\in N(0,1)$,  which must be statistically independent $\mathbb{E}(\tilde{u}_n,\tilde{u}_m)=\delta_{n,m}$. One can thus simulate the stochastic quantum dynamics of a noisy gate very efficiently. Additionally, the  stochastic process is Gaussian and can be  fully described by a simple Lorentzian PSD
\beq
\label{eq:PSD_OU}
    S(\omega)=\frac{ c\tau_{\rm c}^2}{1+(\omega\tau_{\rm c})^2}.
\eeq
We can thus calculate the exact noise-filtered parameters $\Gamma_{1}(t_i),\Delta_1(t_i)$ in Eq.~\eqref{eq:Gammas} for a specific Rabi frequency and pulse duration, and easily calculate the trace distance between the estimated and real noisy gates using either~\eqref{eq:td} or computing $T$ numerically. Alternatively, we can use the random  trajectories generated via Eq.~\eqref{eq:traj_OU} to numerically integrate the stochastic Schr\"{o}inger equation for the qubit  under each imperfect gate in $\mathcal{G}$, and calculate the trace distance with respect to the GST estimates using the numerically averaged density matrices. This method would not be limited by the inherent approximations used to describe the noisy gates using a truncated time-convolutionless master equation (see Appendix~\ref{appendix_2}). In the following, we will report on these two types of benchmarks. {For figures shown in this section, we will use empty markers when benchmarking the first approach, while coloured markers are used for the second approach.} Let us note that in all these benchmarks, we always simulate the GST measurements considering a finite number of shots, such that we account for the finite-sample errors due to shot noise. 

\begin{figure}
  \centering
  \includegraphics[width=1\columnwidth]{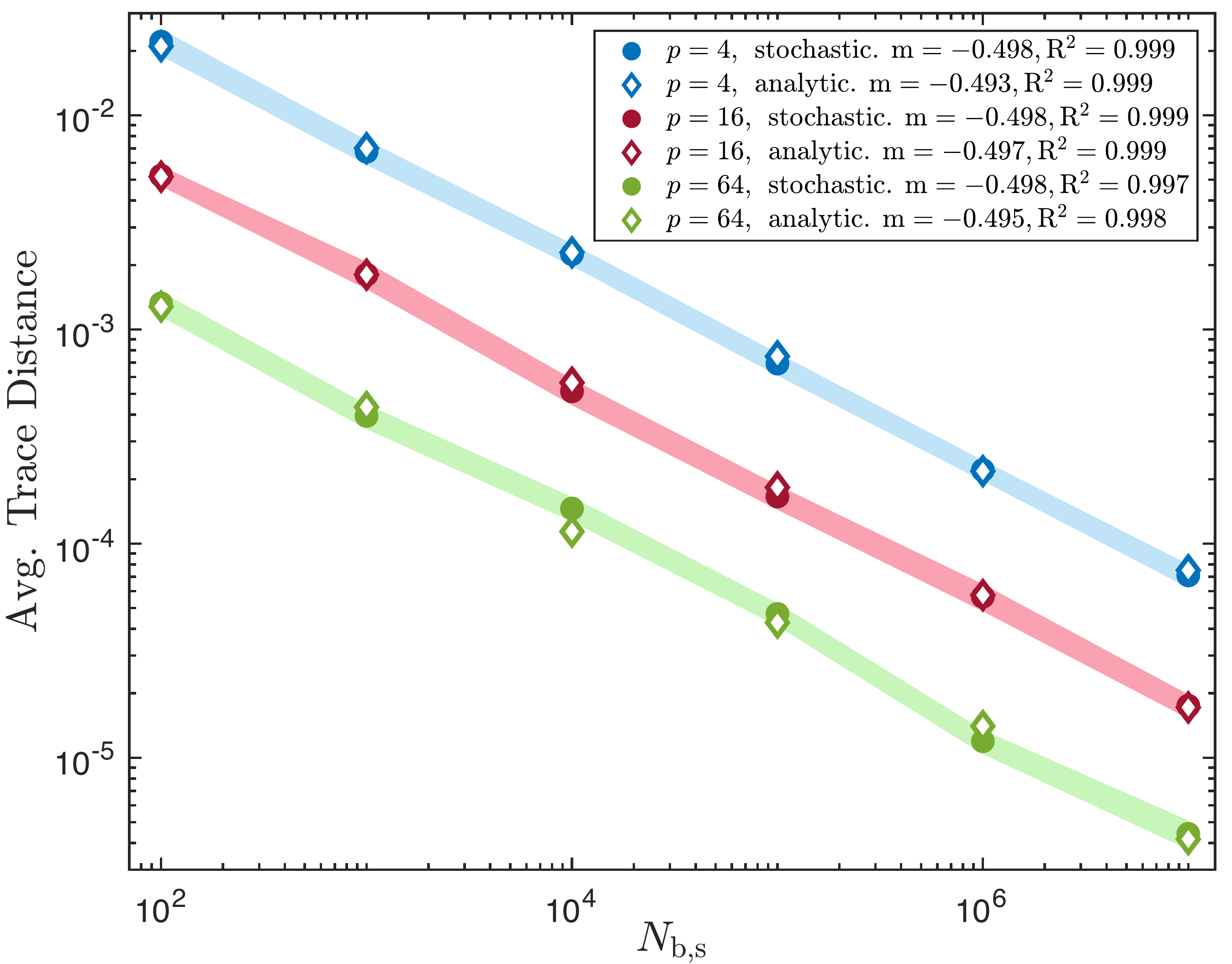}
  \caption{\textbf{${1}/{\sqrt{N_{\rm shots}}}$  scaling with number of measurements:} the average trace distance of the gate set is plotted as a function of the number of shots per circuit for various maximum depths of long-sequence circuits. Estimates for stochastically-generated data are displayed with filled markers, whereas those obtained from the closed filtered-noise expressions for the noisy channel based on the Lorentzian PSD~\eqref{eq:PSD_OU} are displayed with empty markers. Linear fits are carried out for each set of data. For the shake of clearness, we do not plot the linear fits, but the corresponding slopes are shown in the legend, confirming the expected values around $m\simeq-0.5$. For the Ornstein-Uhlenbeck simulations, the noise parameters used are {$\tau_{\rm c}=5\cdot  10^{-6}{\rm s}$, $c=2\cdot  10^3{\rm s}^{-3}$}. 
  To enhance accuracy, we average the process over $100$ GST estimates for each data-point. {In lighter colors, we plot confidence regions with confidence level $\gamma=99\%$.}}
  \label{fig: td_vs_N}
\end{figure}

Let us first present our numerical results for the accuracy of our parametrised GST as a function of the number of measurement shots. As already noted above, we expect  a $1/\sqrt{N_{\rm shots}}$ scaling in the precision of the estimated parameters, which will lead to a similar scaling in  the  trace distance of the estimated gate set with respect to the real microscopic one. We have verified this expected behavior  in Fig.~\ref{fig: td_vs_N}, where we plot the average trace distance versus the number of shots per circuit $N_{ b,s}$. When using a log-log scale, this relationship manifests as a linear function with slope near $-{1}/{2}$, which is very close to  the numerical fits obtained from the numerical GST. 

\begin{figure}
  \centering
  \includegraphics[width=0.94\columnwidth]{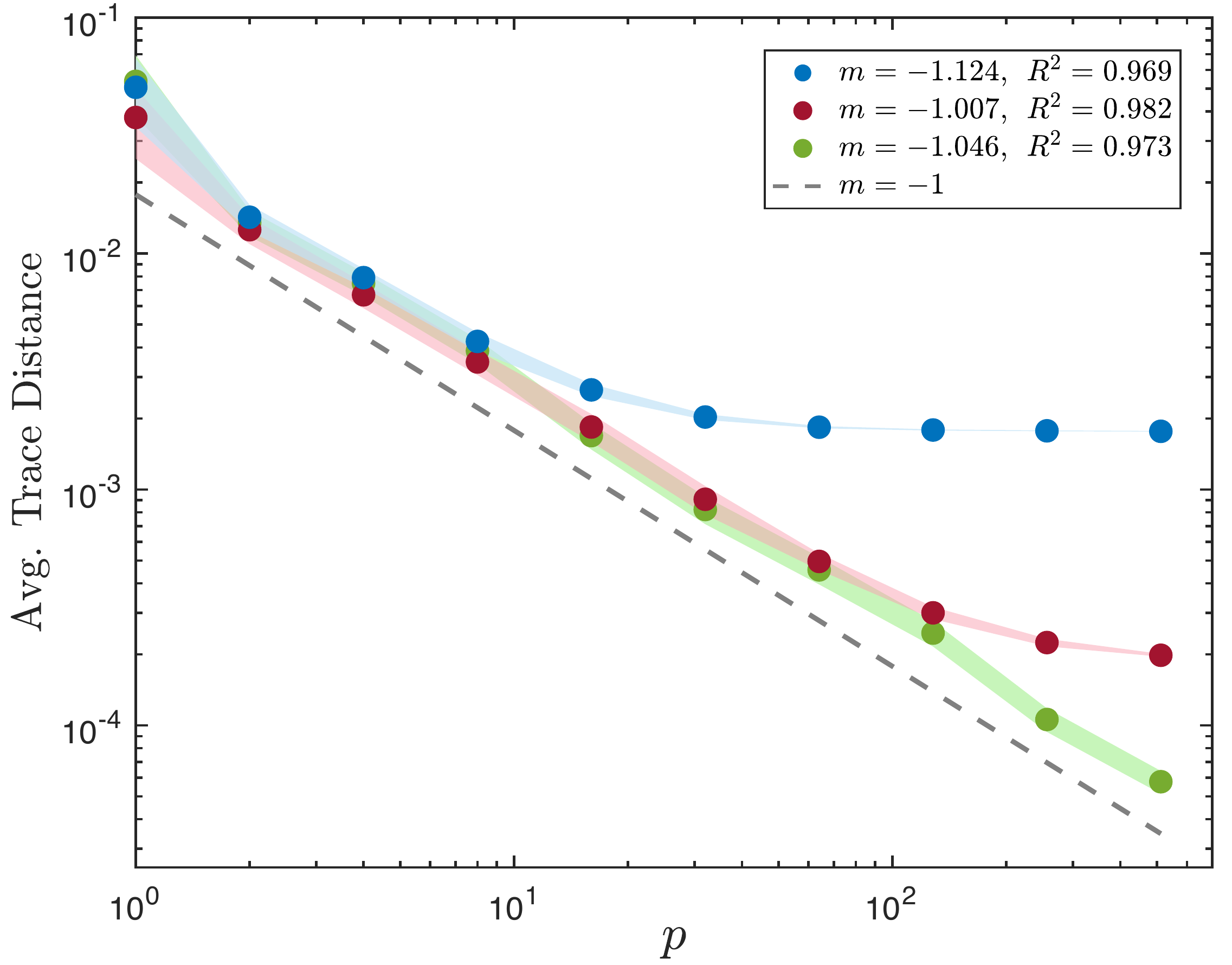}
  \caption{\textbf{${1}/{p}$ scaling  with circuit depth:} {the average trace distance of the gate set is plotted as a function of the maximum number of repetitions of germs utilized in long-sequence circuits for various sets of noise parameters and $N_{b,s}=10^3$ shots per circuit. This process is repeated for $100$ GST estimates at each data-point and then the average values are plotted. Regions in which errors begin to dominate are appreciated. This is in concordance with the $p\gtrsim\frac{1}{\Gamma_1}$ condition anticipated in Sec. \ref{sec_1}. These regions are reflected in the graph by the deviations that magenta and blue data exhibit when compared to the linear function with slope $m=-1$ (dashed grey), whereas green plot matches the tendency for the whole interval displayed. We do not include values within these regions in the linear fits. For the shake of clearness, we do not plot the linear fits, but the corresponding slopes are shown in the legend. These results confirm the expected slope values $m\simeq-1$. Parameters for the Ornstein-Uhlenbeck simulations are $\{
  \tau_{\rm c}=5\times 10^{-4}{\rm s},\hspace{2ex}c=2\times 10^4{\rm s}^{-3}\}$, $\{
  \tau_{\rm c}=5\times10^{-4}{\rm s},\hspace{2ex}c=2\times 10^5{\rm s}^{-3}\}$ and $\{
  \tau_{\rm c}=5\times10^{-4}{\rm s},\hspace{2ex}c=2\times 10^6{\rm s}^{-3}\}$ for the green, magenta and blue sets of data, respectively. In lighter colors, we plot confidence regions with confidence level $\gamma=99\%$.}}
  \label{fig: td_vs_p}
\end{figure}

In a similar manner, we have  also confirmed the expected ${1}/{p}$ scaling for our parametrised long-sequence GST when computing the average trace distance as a function of the maximum number of repetitions of germs utilized in the resulting circuits. This time, the expected slopes on a log-log scale should be around $-1$, which also agrees with our numerical  results depicted in Fig.~\ref{fig: td_vs_p}. Moreover, these results also enable us to confirm the {$p\gtrsim{1}/{\Gamma_1}$} regime in which errors accumulate considerably, and the accuracy of the GST estimation indeed saturates. The deviation from the ${1}/{p}$ behaviour is not observed in the  range of $p$ depths for the green data, as the choice of parameters in this case yields  {$\Gamma_1^{-1}\simeq 10^{5}$. In  contrasts, the saturation becomes noticeable for data in magenta ($\Gamma_1^{-1}\simeq 10^{4}$) and data in blue ($\Gamma_1^{-1}\simeq 10^{3}$)}, where one can see how the initial linear slope bends towards a horizontal line when the accumulation of errors forbids getting more information from going to even longer-depth circuits.

As we previously mentioned, we have focused mostly on the Markovian parametrization of the noisy gates in which $\Gamma_2(t_i)=\Delta_2(t_i)=0$, which is justified when the effects brought up by the non-zero correlation time $\tau_{\rm c}$ are not  too big. As already noted above, our parametrized GST can account for non-Markovian evolutions within each of the gates by simply allowing for $\Gamma_2(t_i),\Delta_2(t_i)\neq0$, but not for possible correlations between different gates. Qualitatively, this requires a separation of time scales such that gates are fast enough for the finite $\tau_{\rm c}$ to play a role, whereas the time lapse between switching off a driving for one gate and switching it on again for the next gate is slow in comparison to  $\tau_{\rm c}$. This captures noise correlations present in a given gate, assuming that correlations are `washed-out' from one gate to the next. 

\begin{figure}
  \centering  \includegraphics[width=1.02\columnwidth]{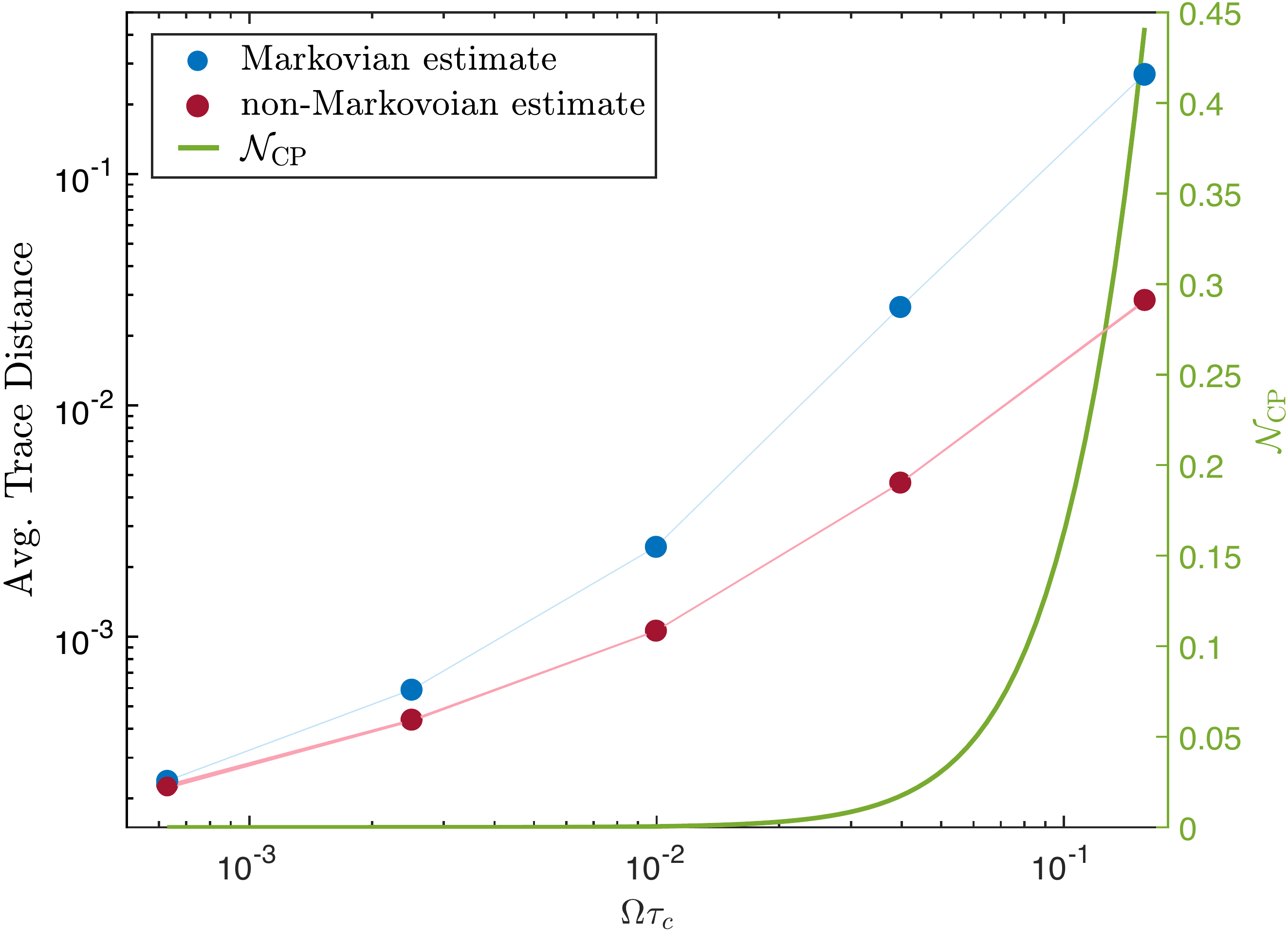}
  \caption{\textbf{Parametrized GST  and non-Markovianity:} {we simulate the non-Markovian GST estimation  by relaxing the approximation $\Gamma_2=\Delta_2=0$. We then run GST on the numerically-simulated data with both a purely Markovian and a non-Markovian  parametrizations. The results from the non-Markovian reconstructions are depicted in magenta, while those from the Markovian reconstructions are shown in blue. The non-Markovian reconstructions effectively capture the short-time correlated effects through the parameters $\Gamma_2$ and $\Delta_2$, thus yielding more accurate tomographic estimates in the regime where $\tau_c$ is larger. 
  In contrast, the Markovian reconstructions fail to capture these effects and exhibit a lower accuracy. In both cases, as $\tau_c$ increases, the condition $\tau_c^3c\ll1$ is relaxed and a worsening of accuracies is observed. We use a secondary Y-axis (green) to plot the non-Markovianity measure $\mathcal{N}_{\rm CP}$ Eq.~(\ref{eq:NM_measure}) in the same correlation times interval. For this graph, all the reconstructions take circuits of a maximum long-sequence length $p=16$ and $N_{b,s}=10^5$. The parameters for the Ornstein-Uhlenbeck simulations employed are
  $\{
  \tau_{\rm c}\in (5\times10^{-7}s,\hspace{1ex}5\times10^{-4}s),\hspace{2ex}c=1.6\times 10^{11}s^{-3}\}$. For each data-point, we average over $100$ GST estimations. Also, in lighter colors, we plot confidence regions with confidence level $\gamma=99\%$.}}
  \label{fig: markovian_vs_non_markovian}
\end{figure}

We can select values of $\tau_{\rm c}$ for which these non-Markovian effects become visible. We account for this effect by comparing the GST estimates obtained without imposing the approximation  $\Gamma_2=\Delta_2=0$. As can be observed in Fig.~\ref{fig: markovian_vs_non_markovian}, forcing $\Gamma_2=\Delta_2=0$ leads to an estimate of a Markovian gate set that becomes less accurate when one reaches a regime in which $\tau_c$ is large enough and  non-Markovian effects start to kick in. By allowing for $\Gamma_2,\Delta_2\neq0$, the parametrized GST can incorporate such non-Markovian effects, leading to gate set estimates that are more accurate. As pointed out in \cite{velazquez2024dynamical}, as one keeps on increasing the correlation time, none of the two parametrizations will be sufficient. In this case, the condition $\tau_c^3c\ll1$ will start to be violated, requiring higher-order terms in the cumulant expansion that underlies the dressed-state master equation and the filtered-noise effective channels. This explains the worsening of both GTS accuracies observed in Fig.~\ref{fig: markovian_vs_non_markovian} as correlation time increases more and more. In any case, we see that the non-Markovian estimate is always better than  the Markovian one. We also use a secondary Y-axis to plot the non-Markovianity measure $\mathcal{N}_{\rm CP}(t=\pi)$ Eq.~(\ref{eq:NM_measure})~\cite{PhysRevLett.105.050403,PhysRevA.89.042120}.

\subsection{Comparison with fully-general GST}

\begin{figure*}[t]
  \centering
  \includegraphics[width=\textwidth]{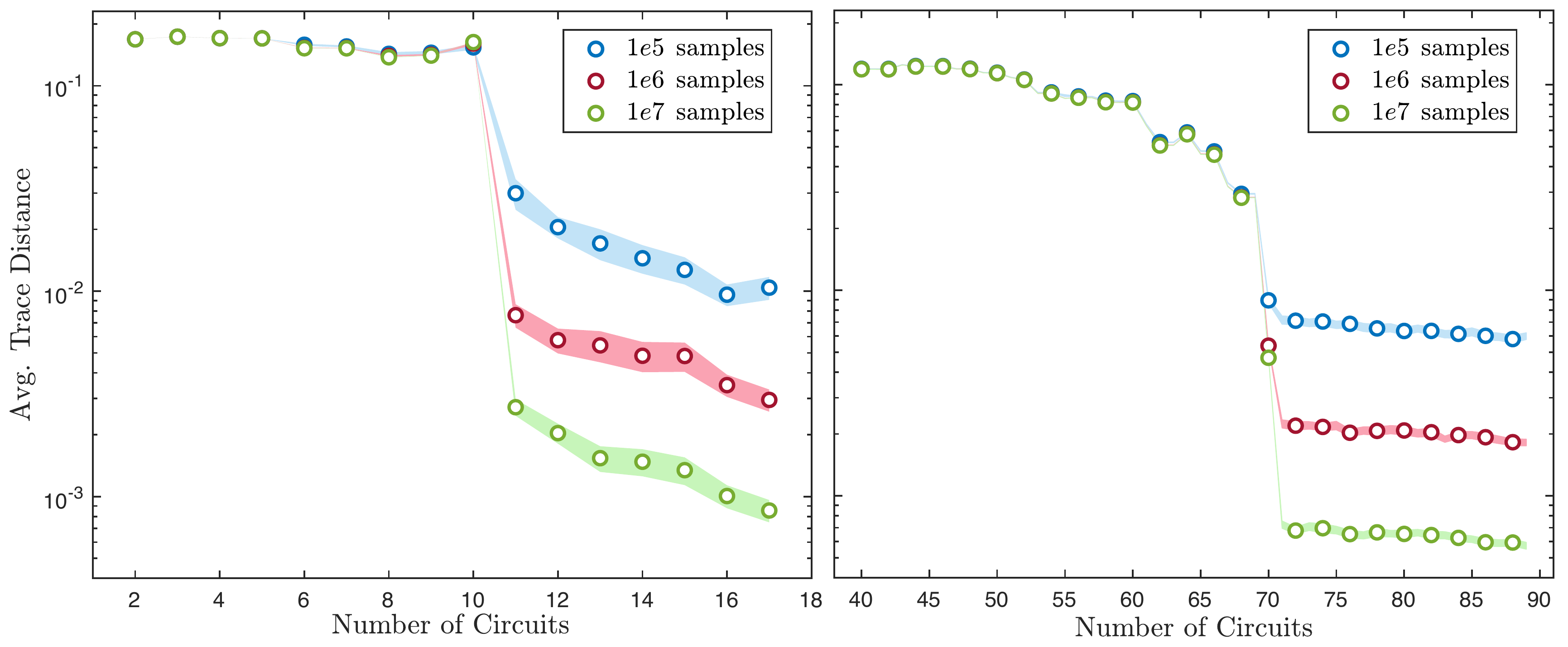}
  \caption{\textbf{Accuracy of GST estimates as a function of the number of circuits included:} we plot the average trace distance of the gate set as a function of the number of circuits included  for both the microscopically-parametrized version of GST (left) and the fully general one (right). In both cases, an improvement in accuracy is observed when the number of non-redundant circuits considered reaches the total number of free parameters in the gate set, which is 67 parameters  for the general model, and 11 parameters for the microscopically-parametrized one. {The  parameters used in these plots are $\tau_{\rm c}=5\times 10^{-4}{\rm s}$ and  $c=2\times 10^9{\rm s}^{-3}$.
  We average the process over $100$ GST estimates for each data-point. In lighter colors, we plot confidence regions with confidence level $\gamma=99\%$.}}
  \label{fig: td_vs_circuits}
\end{figure*}

Once we have checked that our parametrised GST can reproduce the expected scalings in accuracy, and that it can also  account for a certain amount of non-Markovian effects, we can finally compare our estimates  with those obtained utilizing fully-general GST. As we mentioned in Sec. \ref{sec_1}, constructing a minimal GST design requires significantly fewer experiments when using our GST based on a reduced set of noise-filtered parameters. This motivates the study presented in Fig.~\ref{fig: td_vs_circuits},  where we plot  the average trace distance between the estimated and microscopic gate  set as a function of the number of  circuits $N_{\rm c}$ used in the GST estimations. In this case, we assume  $p=1$ for simplicity, and include the SPAM elements when computing the average trace distance to account for every parameter in the gate set. This allows us to compare the behaviour of  parametrized and fully-general GST and, particularly, how the accuracy of the estimation changes as we  increase the number of base circuits used in GST. 

When the number of circuits is not sufficient to reach  IC, Fig.~\ref{fig: td_vs_circuits} shows that the GST estimation does not work, as the trace distance remains close to that  of the initial unitary guess and the actual noisy channels. Since the GST sensitivity to the parameters is very low in this regime of a low-number of circuits,   the estimation cannot learn the noisy parameters correctly. As we increase the number of circuits and  IC  is attained, an abrupt drop in the average trace distance to the actual gate set is observed, signalling that the GST is now accurate. This can be attributed to the fact that every parameter in the gate set is now being optimized when performing the gradient descent of the maximum-likelihood estimation. 

We note that the number of circuits at which this drop occurs coincides, for our parametrized model, with the number of free parameters of the noisy gate set $N_{GST}=11$ (left panel of Fig.~\ref{fig: td_vs_circuits}). This close agreement  is a result  of the circuit selection procedure described in subsection \ref{circuit_selection}, which enables us to order circuits according to their sensitivity, and place the first 11 non-redundant elements at the beginning. In contrast, for fully-general GST (right panel of Fig.~\ref{fig: td_vs_circuits}), we observe that the drop in trace distance is only reached further along the circuit axis, requiring thus a larger number of circuits. This should be unsurprising, as the fully-general GST   requires estimating  $N_{GST}=67$  parameters. In contrast to our parametrized GST,  where the circuit importance sampling is direct, we rely on the \texttt{pyGSTi}  algorithms for fully-general GST to find the minimal set of circuits sufficient to achieve information completeness, according to which $72$ different experiments are needed. This number coincides exactly with the point in which trace distance drops in the right panel of Fig.~\ref{fig: td_vs_circuits}.

\begin{figure}[ht]
  \centering
  \includegraphics[width=1\columnwidth]{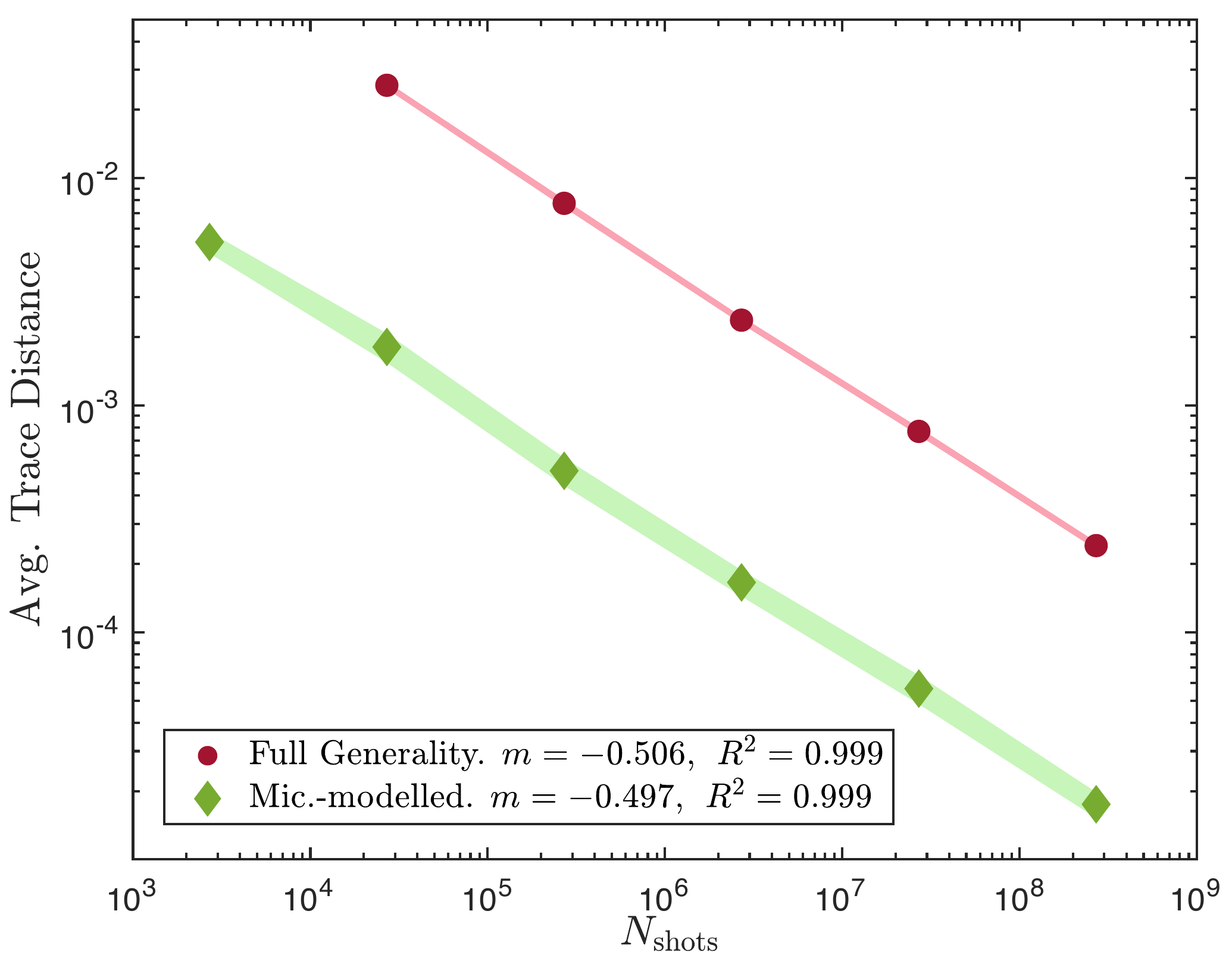}
  \caption{\textbf{Comparison with fully general GST:} we present the average trace distance of the gate set as a function of the total number of shots used $N_{\rm shots}$. In the plot, our microscopically-motivated implementation is depicted in green with diamond markers, while fully general GST implementations is depicted in magenta. It is notable that our parametrized GST  achieves a better estimation accuracy even with a significantly reduced total number of circuits required. As expected, accuracy scales as $1/\sqrt{N_{\rm shots}}$ in both cases, as evident from the linear fit slopes. The estimate obtained for $N_{\rm shots}\simeq10^3$ is only displayed for our parametrized approach, as the number of shots is insufficient to perform GST with the fully-general one . Each plot is generated using $p=16$ and the simulation parameters {$\{
  \tau_{\rm c}=5\times 10^{-6}{\rm s},\hspace{2ex}c=2\times 10^3{\rm s}^{-3}\}$. We also average the results over $100$ iterations to limit the variance and we plot confidence regions with confidence level $\gamma=99\%$ in lighter colors.}}
  \label{fig:comparison}
\end{figure}

To give estimates of the total measurement resources   required for  GST, and compare our parametrized approach to the fully-general one, we finally analyze the average trace distance of the gate set as a function of the total number of shots $N_{\rm shots}$. As expected, our parametrized GST provides a significant reduction in the required number of experiments to achieve a target accuracy in comparison  to fully-general GST (see  Fig.~\ref{fig:comparison}). We  observe that the accuracy of our parametrized GST  outperforms that of fully-general GST by more than an order of magnitude when considering the same total number of shots $N_{\rm shots}$. This can be seen  when comparing the green data (parametrized GST) with the magenta data (fully-general GST) in Fig.~\ref{fig:comparison}. Note that in this figure we are fixing the same maximum depth for the circuits in the long sequence scheme with both models. Thus, the total number of shots $N_{\rm shots}$ grows by increasing the number of shots per circuit $N_{\rm b,s}$ employed to compute each frequency $f_{\mu,s}$.


\begin{figure}
  \centering
  \includegraphics[width=1\columnwidth]{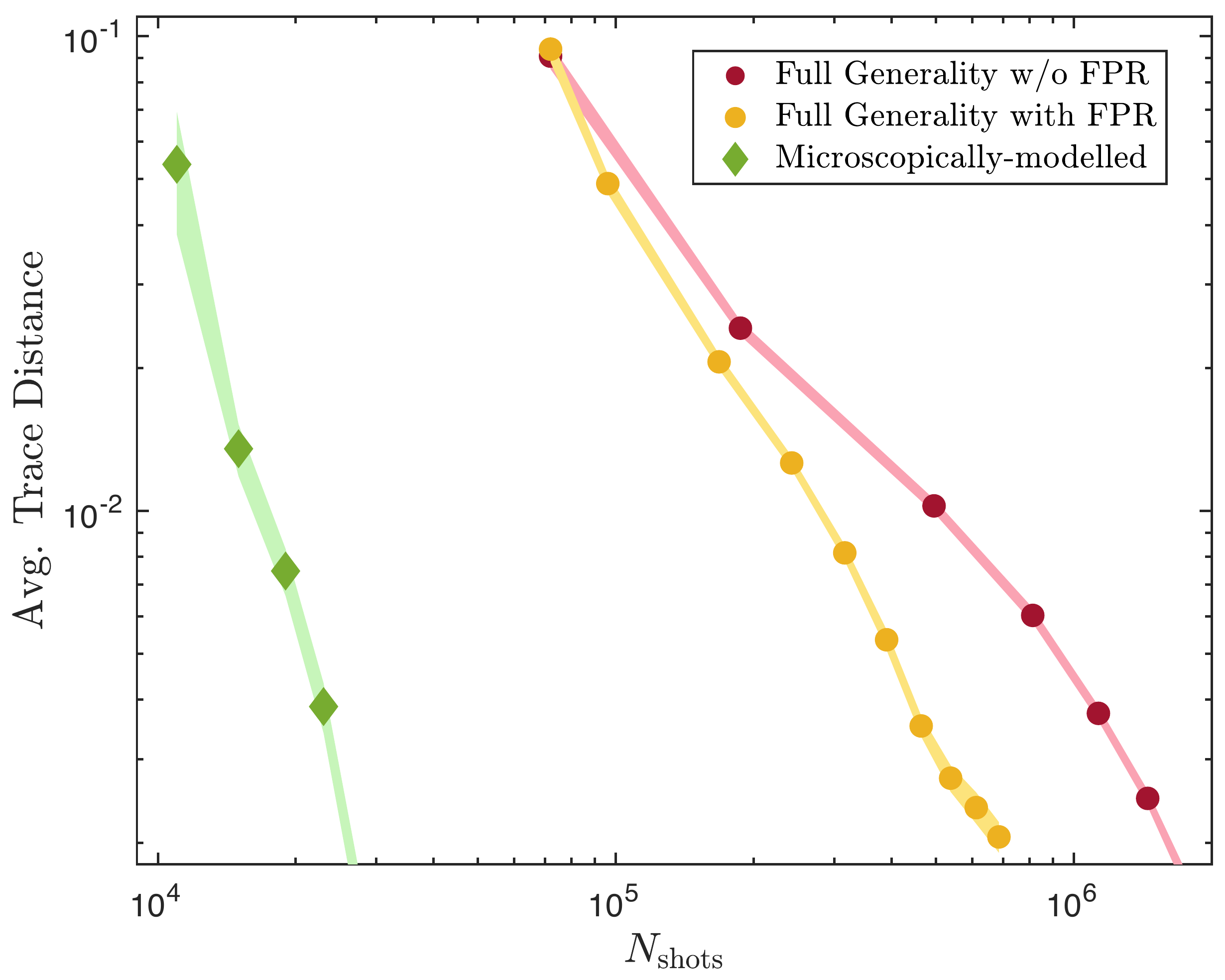}
  \caption{\textbf{Comparison including fiducial pair reduction in fully general GST:} we present the average trace distance of the gate set as a function of the total number of shots used $N_{\rm shots}$. In the plot, our microscopically-parametrized GST is depicted in green with diamond markers, while fully general GST implementations are shown in other colors. Specifically, the yellow and magenta plots represent general GST estimations with and without FPR, respectively. Contrary to Fig.~\ref{fig:comparison}, we fix the number of shots per circuit, in this case is $N_{ b,s}=10^3$. The total number of shots $N_{\rm Shots}$ varies by considering increasingly deeper circuits, ranging in the interval $p_g\in\{2^i\}_{g=0}^{9}$. It is notable that our parametrized  approach achieves much better estimation accuracies at a fixed $N_{\rm Shots}$. Additionally, it is worth mentioning that the improved resource allocation stemming from FPR does also improve the estimations, as evident from the yellow plot. Each plot is generated using the simulation parameters {$\{
  \tau_{\rm c}=5\times 10^{-4}{\rm s},\hspace{2ex}c=2\times 10^4{\rm s}^{-3}\}$. We also average the results over $100$ iterations to limit the variance and plot confidence regions with confidence level $\gamma=99\%$ in lighter colors.}}
  \label{fig:comparison_optimal}
\end{figure}

Let us now discuss how the accuracy of the estimation presented in Fig.~\ref{fig:comparison} can  be further enhanced. Similar to that figure, Fig.~\ref{fig:comparison_optimal} illustrates the average trace distance as a function of $N_{\rm shots}$. However, in contrast to Fig.~\ref{fig:comparison}, we now more wisely choose to increase $N_{\rm shots}$ by gradually incorporating deeper circuits in the GST scheme while maintaining the same number of shots $N_{ b,s}$. This accelerates the improvement in accuracy with the number of shots, as already discussed and evident from Figs.~\ref{fig: td_vs_N} and~\ref{fig: td_vs_p}. This aligns with the logic behind the utility of FPR. The fact that one is able to reduce the number of circuits needed to achieve an AC set enables employing the excess of resources not to increase $N_{ b,s}$ (which would generally yield similar results to those without FPR), but to extend circuits further in depth. Consequently, with this expanded range of depths, FPR provides superior results at a constant number of shots. This can be observed when comparing the magenta (no FPR) and the yellow (FPR) curves in Fig.~\ref{fig:comparison_optimal}. Likewise, our microscopic parametrization does also reduce the number of circuits needed to achieve an AC set. Therefore, akin to FPR, we can allocate the resources more efficiently among deeper circuits. The corresponding plot in Fig.~\ref{fig:comparison_optimal} is displayed in green. As can be concluded from this figure, the accuracy enhancement obtained with our parametrized  approach is great when proceeding in this way.

\subsection{Inclusion of amplitude noise fluctuations}
\label{amplitude noise}

As extensively detailed in previous sections of this manuscript, our work primarily aims at describing processors subjected to phase noise using the filter-function formalism. Nonetheless, this model can be further extended to additionally account for amplitude or `intensity' fluctuations of the drive. As pointed out in Appendix~\ref{appendix_1}, these additional fluctuations are similarly modelled by an stochastic process $\delta\tilde{\Omega}(t)$, which describes the variations of the Rabi frequency induced by intensity noise~\cite{Zoller_1978}. Equivalently to the dressed-state master equation description of the pure phase noise evolution, amplitude noise leads to a supplementary filtered integral analogous to those in Eq.~(\ref{eq:Gammas})
\begin{equation}\label{eq: amplitude_correction}
    \Delta\Gamma_1(t) = \int_{-\infty}^\infty \!\!\!{\rm d}\omega {S}_{\Omega}(\omega)F_\Omega(\omega,t),
\end{equation}
where we have introduced an additional PSD, ${S}_{\Omega}$, which exclusively describes this process. The new filter function $F_\Omega(\omega,t)$ is explicitly given in Eq.(~\ref{eq:filterDeltaGamma}).

\begin{figure}
  \centering
  \includegraphics[width=1\columnwidth]{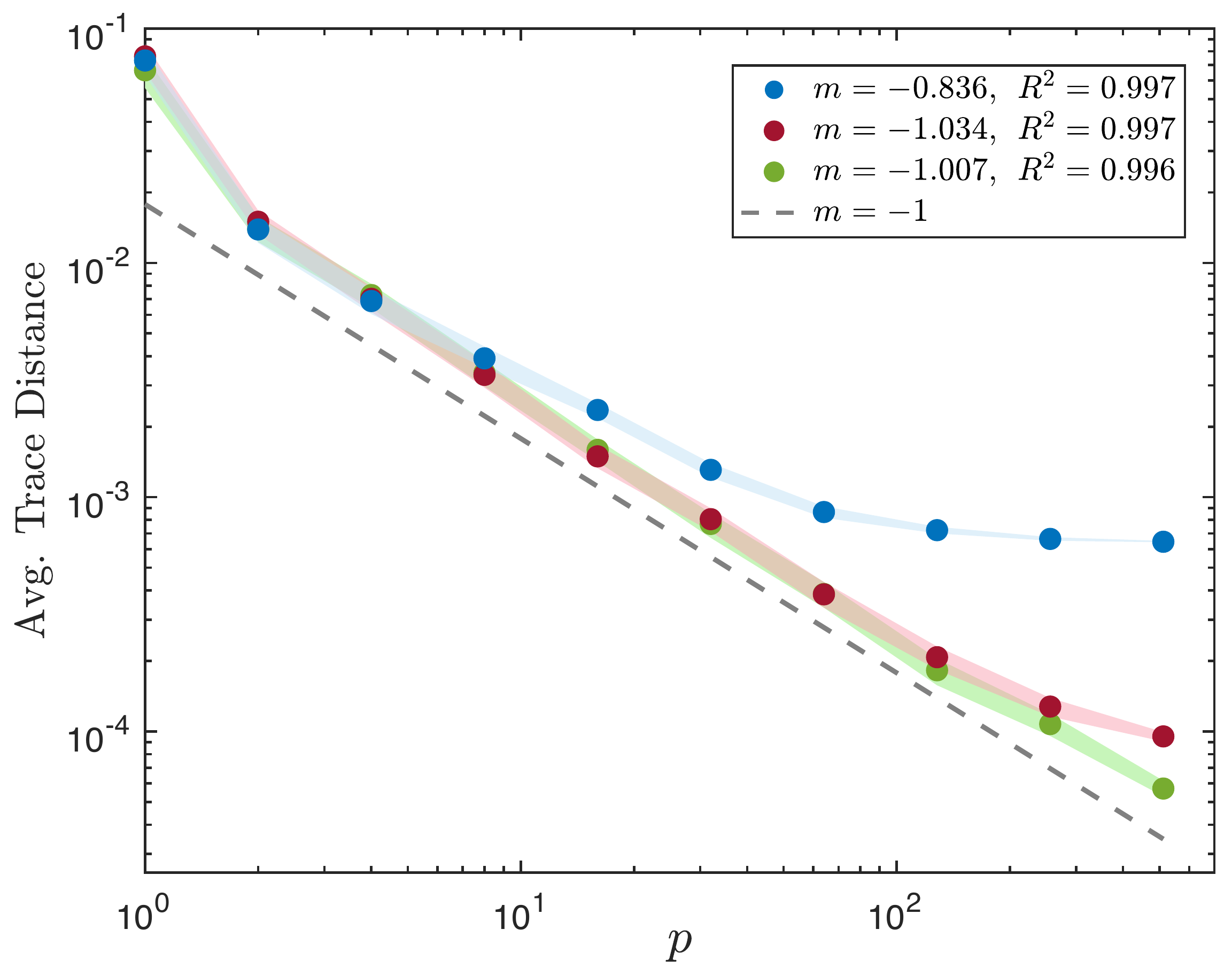}
  \caption{\textbf{${1}/{p}$ scaling  with circuit depth for phase and amplitude fluctuations:} {the average trace distance of the gate set is plotted as a function of the maximum number of repetitions of germs utilized in long-sequence circuits for various sets of noise parameters and $N_{b,s}=10^3$ shots per circuit. This process is repeated for $100$ GST estimates at each data-point and then the average values are plotted. Regions in which errors begin to dominate are appreciated. This is in concordance with the $p\gtrsim\frac{1}{\Gamma_1}$ condition anticipated in Sec. \ref{sec_1} under the substitution $\Gamma_1\to\Gamma_1+\Delta\Gamma_1$. These regions are reflected in the graph by the deviations that magenta and blue data exhibit when compared to the linear function with slope $m=-1$ (dashed grey), whereas green plot matches the tendency for the whole interval displayed. We do not include values within these regions in the linear fits, neither the first data-point, which regularly encounters a local minima. For the shake of clearness, we do not plot the linear fits, but the corresponding slopes are shown in the legend. These results confirm the expected slope values $m\simeq-1$. Parameters for the Ornstein-Uhlenbeck simulations are $\{
  \tau_{\rm c}=5\times 10^{-4}{\rm s},\hspace{2ex}c=2\times 10^4{\rm s}^{-3}\}$, $\{
  \tau_{\rm c}=5\times10^{-4}{\rm s},\hspace{2ex}c=2\times 10^5{\rm s}^{-3}\}$ and $\{
  \tau_{\rm c}=5\times10^{-4}{\rm s},\hspace{2ex}c=2\times 10^7{\rm s}^{-3}\}$ for the amplitude fluctuations in the green, magenta and blue sets of data, respectively. For every simulation, we employ $\{
  \tau_{\rm c}=5\times 10^{-4}{\rm s},\hspace{2ex}c=2\times 10^4{\rm s}^{-3}\}$ in the phase noise stochastic processes. We recall that phase and amplitude stochastic fluctuations are not cross-correlated.  In lighter colors, we plot confidence regions with confidence level $\gamma=99\%$.}}
  \label{fig: td_vs_p_intensity}
\end{figure}

\begin{figure}
  \centering
  \includegraphics[width=1\columnwidth]{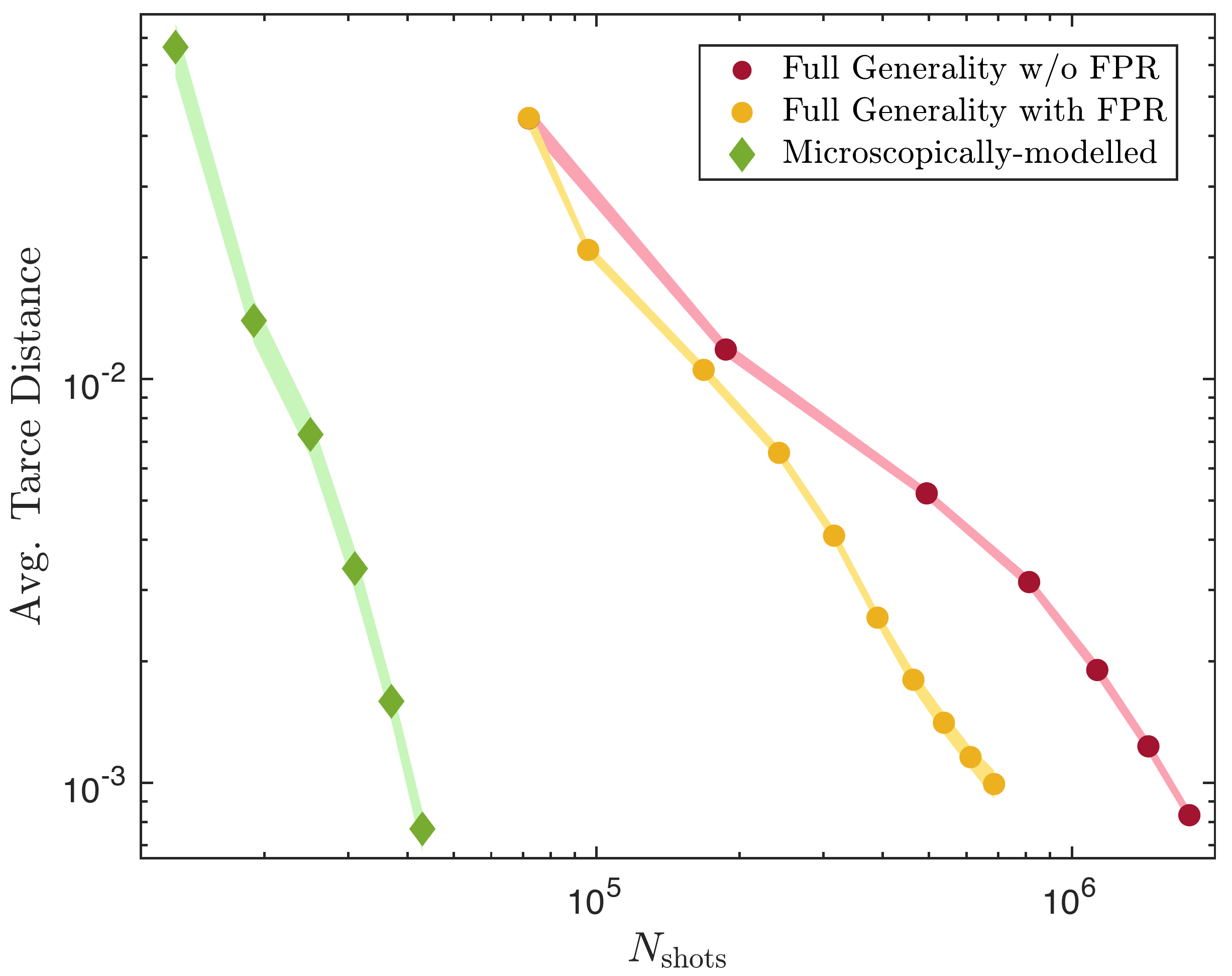}
  \caption{\textbf{Comparison including fiducial pair reduction in fully general GST and amplitude fluctuations:} we present the average trace distance of the gate set as a function of the total number of shots used $N_{\rm shots}$. In the plot, our microscopically-parametrized GST is depicted in green with diamond markers, while fully general GST implementations are shown in other colors. Specifically, the yellow and magenta plots represent general GST estimations with and without FPR, respectively. Contrary to Fig.~\ref{fig:comparison}, we fix the number of shots per circuit, in this case is $N_{ b,s}=10^3$. The total number of shots $N_{\rm Shots}$ varies by considering increasingly deeper circuits, ranging in the interval $p_g\in\{2^i\}_{g=0}^{9}$. It is notable that our parametrized  approach achieves much better estimation accuracies at a fixed $N_{\rm Shots}$. Additionally, it is worth mentioning that the improved resource allocation stemming from FPR does also improve the estimations, as evident from the yellow plot. Each plot is generated using the simulation parameters {$\{
  \tau_{\rm c}=5\times 10^{-4}{\rm s},\hspace{2ex}c=2\times 10^4{\rm s}^{-3}\}$ for both phase and amplitude fluctuations. We recall that these two stochastic processes are not cross-correlated. We also average the results over $100$ iterations to limit the variance and plot confidence regions with confidence level $\gamma=99\%$ in lighter colors.}}
  \label{fig:comparison_optimal_intensity}
\end{figure}

As can be anticipated from Eq.~(\ref{eq: amplitude_correction}), the parameter $\Delta\Gamma_1(t)$ neither depends on the driving phases $\phi_i$. Thus, the fact that gates with shared pulse duration are described by the same set of parameters prevails. Consistently with our previous analysis, we employ Fig.~\ref{fig: td_vs_p_intensity} as a confirmation that depth-one germs still define an AC set. For this figure, as well as in Fig.~\ref{fig:comparison_optimal_intensity}, we simulate data using two independent stochastic processes under OU niose, i.e. they do not exhibit cross correlations (see Apendix~\ref{appendix_1}). As usual, we take an average over a sufficiently large number of random trajectories. It is noteworthy that we do not observe an emerging gauge degree of freedom induced by the parameter $\Delta\Gamma_1(t)$, therefore, the need of gauge optimization remains absent with this extended model. Using this knowledge and taking advantage of \texttt{pyGSTi} adaptability, we can effortlessly obtain a new figure equivalent to Fig.~\ref{fig:comparison_optimal} which does also include amplitude noise fluctuations. As can be expected, the inclusion of two additional parameters to fully characterize the gate set under phase and amplitude noise does not significantly alter the behaviour of the results (see Fig.~\ref{fig:comparison_optimal_intensity}). As a consequence, the discussion concerning Fig.~\ref{fig:comparison_optimal} remains accurate.

\section{\bf{Conclusions and outlook}}
\label{conclusions}
In this work, we have presented a microscopically-motivated   GST, which builds on the quantum dynamical map of single-qubit gates subject to phase noise~\cite{velazquez2024dynamical} to find a compact parametrization of the full gate set in terms of filtered integrals of the noise spectral density. We have generalised this work to parametrise the full gate set, and discussed central differences with respect to the standard fully-general GST, such as  the possibility of using single gates as base circuits to achieve amplification completeness. In addition, we showed that germ selection, fiducial pair reduction, and thus the achievement of minimal GST designs can be highly simplified when using this microscopic parametrization. We also highlighted how our approach does not suffer from gauge redundancy,   saving  computational resources as no gauge-fixing is required, and avoiding reference frame issues when benchmarking the results. We have presented a thorough  numerical validation of these arguments considering a specific stochastic model of coloured phase noise, and numerically obtaining maximum-likelihood estimates based on long-sequence GST with numerically-generated  data that also includes the effects of shot noise. We have shown that the accuracy of our parametrized long-sequence GST follows the expected scalings with the number of shots and the circuit depth, and demonstrated that our parametrization can account for non-Markovian dynamics during the gates. We have finally showcased the advantages of our parametrized GST with respect to fully-general GST, presenting examples in which a tenfold reduction in the average trace distance can be achieved considering the same number of total measurement shots and circuit depth. We also showed how this reduction can be increased if one allocates the remaining resources in experiments with  deeper circuits. Additionally, we showed that these results remain accurate under an extension of the model to include amplitude noise fluctuations.

As an outlook, it would be interesting if future studies  assessed if these benefits  remain valid when extending to two-qubit gates and crosstalk in an enlarged qubit register, as required for realizing the full potential of quantum computation. On the one hand, the fact that single gates themselves form an amplificationally complete set is something dependent on the specific matrix structure of our  parametrized model. An equivalent treatment for entangling gates should be developed to check whether more complex germs are needed for amplification completeness in that case. In order to do that, an extension of the effective quantum dynamical map to model noisy two-qubit gates will be required. Once this is made, we believe that fiducial pair reduction would not longer yield minimal but near-minimal GST designs. We note here that adding entangling gates will still generate disjoint sets of per-germ amplified parameters if the two-qubit gates do also serve as germs. This would facilitate finding good choices for fiducial pairs, but the $d-1$ independent outcomes of each circuit would still lead to some overhead if each parameter is linked to a single experimental setting, resulting in a more complex combinatorial problem. On the other hand, it could also be the case that some of the new parameters describing the entangling gates are gauge parameters, while the single-qubit gate parameters would remain non-gauge ones. Even in the worst-case scenario in which none of the previous features persists in the two-qubit case, the considerable reduction in the total number of parameters still makes it worth to explore this research avenue. Considering  entangling gates in a two-qubit system and also  cross-talk in the previous single-qubit gates, the number of parameters needed to characterize this gate set with fully-general GST grows to $N_{GST}=5103$. On the contrary, even with no further modelling of the entangling gate, using the present  physically-motivated parametrization yields $N_{GST}=423$ or $N_{GST}=503$ for the Markovian and the non-Markovian estimations respectively, which will be further reduced by a detailed modelling of the noisy entangling gate.


Another interesting question for future research will involve validating our approach with real experimental data,  assessing the extent to which our analytical model accurately describes physical quantum devices. The use of this model for trapped-ion QIPs is justified in \cite{velazquez2024dynamical} by the fidelity agreement when comparing tomographic results with analytical estimations, but the model can be ultimately extended to include other sources of error, such as intensity noise and spontaneous decay. Once a description of the faulty processes of a device is obtained, it can be further tested using techniques like probabilistic error cancellation (PEC) \cite{Temme_2017}, which allows to obtain ideal expectation values in terms of the accessible faulty ones.

Finally, we also note that it would be interesting  to extend the description of non-Markovian effects briefly introduced in the final part of subsection~\ref{num_results} beyond the individual gates, such that non-Markovian correlations among  the different gates that form the base circuits can be accounted for, which would lead to a fully non-Markovian generalization of GST.

\acknowledgements
The authors warmly thank J.M. Sánchez Velázquez for his enlightening discussions and assistance on countless occasions. Additionally, both P.V. and A.B.  thank Ch. D. Marciniak and Claire L. Edmunts for discussions during the development of this work.
Finally, P.V. would like to thank Stefan Seritan and the team behind \texttt{pyGSTi} for their technical support on the use of the GST python package.

The project leading to this publication has received funding from the US Army Research Office through Grant No. W911NF-21-1-0007. A.B acknowledges support from PID2021-127726NB- I00 (MCIU/AEI/FEDER, UE), from the Grant IFT Centro de Excelencia Severo Ochoa CEX2020-001007-S, funded by MCIN/AEI/10.13039/501100011033, from the CSIC Research Platform on Quantum Technologies PTI-001, and from the European Union’s Horizon Europe research and innovation programme under grant agreement No 101114305 (“MILLENION-SGA1” EU Project). Views and opinions expressed are, however, those of the author(s) only and do not necessarily reflect those of the European Union or the European Commission. Neither the European Union nor the granting authority can be held responsible for them.



\appendix

\section{ Quantum tomography and GST}
\label{appendix_2}
Quantum Tomography techniques are grounded in the ability to infer a representation of a quantum state, a quantum channel, or a set of them from  experimental data. In this Appendix, for the sake of completeness, we review standard tomographic approaches, which  serves to set our notation. 

When the aim is to estimate a quantum state $\rho\in\mathsf{D}(\mathcal{H})$, a set of predetermined {positive operator valued measure (POVM)} elements  $\{M_\mu:\,\mu\in\mathbb{M}\}$ must be specified. This set must satisfy $M_\mu\in\mathsf{Pos}(\mathcal{H})$ and $\sum_\mu M_\mu=\mathbb{1}_2$, and should also form a basis of the $d$-dimensional Hilbert space sufficient to uniquely identify the state of interest. The POVM meeting this criteria is said to define an {informationally complete} (IC) set, and should comprise at least $d^2$ elements, being overcomplete when more than $d^2$ elements are included in the set. In this case, the state can be estimated from the measurement probabilities
\begin{equation}
\label{eq: prob_Appendix}
p_{\mu}=\mathrm{Tr}\left\{M_{\mu}\rho\right\}.
\end{equation}

For the ongoing analysis, it will prove convenient to define the {super-operator} formalism \cite{kitaev2002classical}, in which a density matrix and a POVM element $\rho,M_\mu$ are written as vectors $|\rho\rangle\!\rangle,|M_\mu\rangle\!\rangle $ whose elements are the coefficients associated to its decomposition in a Hilbert space basis. Quantum channels  appear as matrices acting on these vectors, such that  $\mathrm{Tr}\left\{M_{\mu}\mathcal{E}(\rho)\right\}=\langle\!\langle M_{\mu}|G_{\mathcal{E}}|\rho\rangle\!\rangle$, where $G_{\mathcal{E}}$ denotes the super-operator corresponding to $\mathcal{E}$. When the Pauli basis is employed, this is commonly referred to as the {Pauli transfer matrix} (PTM) representation~\cite{greenbaum2015introduction}, in which the elements of $G_{\mathcal{E}}$ are expressed as 
\begin{equation}
[G_{\mathcal{E}}]_{\alpha,\beta}=\mathrm{Tr}\left\{E_\alpha\mathcal{E}(E_\beta)\right\}.
\end{equation}

Returning to quantum state tomography (QST), using the experimental frequencies as an estimate of the probabilities $f_{\mu}\simeq p_{\mu}$, Eq.~(\ref{eq: prob_Appendix}) trivially leads to the expression
\begin{equation}
\label{eq: linear_relation}
|f\rangle\!\rangle = \begin{pmatrix} f_1,f_2,\cdots,f_{d^2}\end{pmatrix}^{\!T}= A|\rho\rangle\!\rangle,
\end{equation}
where the matrix $A$ is built using the vectorized POVM elements as follows
\begin{equation}
A = \begin{pmatrix} |M_1\rangle\!\rangle, | M_2\rangle\!\rangle,\cdots,| M_{d^2}\rangle\!\rangle\end{pmatrix}^{\!T}.   
\end{equation}
It is then straightforward to invert Eq.~(\ref{eq: linear_relation}) to obtain the estimated state $\hat{\rho}$, namely
\begin{equation}
|\hat{\rho}\rangle\!\rangle = (A^TA)^{-1}A^T|f\rangle\!\rangle,
\end{equation}
where we have assumed that $A$ is not generally a square matrix.
A schematic representation of QST is shown in Fig.~\ref{fig: tomography_scheme}\textcolor{magenta}{a} at the introduction.

When the objective is to reconstruct a quantum process $\mathcal{E}$ rather than a state, an IC set of predetermined initial states $\{\rho_s,s\in\mathbb{S}\}$ has to be specified in addition to $\{M_{\mu},\mu\in\mathbb{M}\}$. This tomographic technique is known as quantum process tomography (QPT), and the measurements required are now obtained from the action of $\mathcal{E}$ upon each initial state. These `sandwiches' $p_{\mu s} = \langle\!\langle M_{\mu}|G_{\mathcal{E}}|\rho_s\rangle\!\rangle$ can be visualized in Fig.\ref{fig: tomography_scheme}\textcolor{magenta}{b}. Once the experimental frequencies $f_{\mu s}$ have been obtained, one can  use linear inversion to get the super-operator estimate
\begin{equation}
\label{eq: linear_QPT}
\hat{G}_{\mathcal{E}}=(A^TA)^{-1}A^TFB^T(BB^T)^{-1},
\end{equation}
where $(F)_{\mu, s}=f_{\mu s}$ and $B=(|\rho_1\rangle\!\rangle,  |\rho_2\rangle\!\rangle,\cdots, |\rho_{d^2}\rangle\!\rangle)$.

In practice, owing to finite-sample shot noise  and other experimental imperfections, linear inversion can lead   to estimated processes  that are not physical $\hat{G}_{\mathcal{E}}\mapsto\hat{\mathcal{E}}\notin\mathsf{C}(\mathcal{H})$, in the sense that they do not meet the complete-positivity and trace preservation (CPTP) constraints. This motivates the use of maximum likelihood  estimation (MLE), which allows for further constraints on the process. This method relies on the observation that, with proper normalization of the probabilities, the multinomial distribution
\begin{equation}
\label{eq:MLE_QPT}
\Lambda(\mathcal{E})=\prod\limits_{\mu, s}\left(p_{\mu s}(\mathcal{E})\right)^{f_{\mu s}},    
\end{equation}
maximizes precisely when $p_{\mu s}(\mathcal{E})=f_{\mu s}$, where $p_{\mu s}(\mathcal{E})$ depend on the parametrized channel. Hence, by solving for the maximum likelihood, one obtains an estimate $\hat{\mathcal{E}}$ that is both physical and shows the highest likelihood to reproduce the observations. Alternatively, one can  define a cost function via the negative log-likelihood  $\mathcal{L}=-\log\Lambda(\mathcal{E})$, and minimize it $\hat{\mathcal{E}}=\texttt{argmin}\{\mathcal{L}\}$ subject to the physically-admissible constrains through which  CPTP is enforced. It is also common to use a least-squares cost function $\mathcal{C}$ as a proxy of $\mathcal{L}$, albeit behaving as a biased estimator for rare events \cite{Nielsen_2021}. 
In the PTM representation $\hat{\mathcal{E}}\mapsto\hat{G}_{\mathcal{E}}$, TP translates into $[\hat{G}_{\mathcal{E}}]_{1,j}=\delta_{1,j}$, such that only the diagonal element of the first row differs from zero. Unfortunately, CP does not have a direct effect in this PTM representation. On the other hand, as noted  in Sec.~\ref{sec_1} of the main text, both CP and TP are readily imposed using the process matrix representation. If $\chi$ denotes the process matrix of the quantum channel of interest~\eqref{eq:chi_matrix}, CPTP constraints can be expressed as
a positive semidefinite condition    and a so-called~\cite{Chuang1997} trace constraint 
 \begin{equation}
 \label{eq:trace_constraint}
 \chi\in\mathsf{Pos}(\mathbb{C}^{d^2}),\hspace{1ex} \sum_{\alpha, \beta} \chi_{\alpha, \beta} E_{\beta}^{\dagger}E_{\alpha}^{\phantom{\dagger}}=\mathbb{1}_{d^2}.
\end{equation}

As outlined during the introduction, QPT encounters self-consistency issues, particularly when specifying the IC sets. These sets are typically obtained by applying faulty operations to a fiducial state and  measurement which, although can be  prepared reliably in the experiment, are also subject to imperfections. It is often the case that the gate one aims at characterizing is also needed to prepare the IC set, leading to an inherent inconsistency. Addressing this challenge was the catalyst for gate set tomography (GST), which we now  outline.

GST aims to go beyond the characterization of an individual quantum channel, targeting a full gate set $\mathcal{G}$ that is composed of   different elements. These elements are generally referred to as {gates}, and  should  enable for the construction of IC sets at SPAM stages, achieved  by acting on fiducial operators with either a set of imperfect individual   gates $\{G_i:\, i\in\mathbb{G}\}$ or combinations thereof, here denoted by $\{H_s,H_b:\, s\in\mathbb{S},b\in\mathbb{M}_b\}$. In addition to these imperfect SPAM gates, GST also includes  
the fiducial initial state(s)  and projective measurement(s) in the gate set $\mathcal{G}$. To keep notation simple, we assume there is only one of each of these native fiducial elements $\rho_0,M_{0}$, which holds for most experimental setups such as $N=1$ where $\rho_0\approx\ket{0}\!\bra{0}$, and $M_{0}\approx(1+\sigma_z)/2$. We note that there are direct extensions of this situation that include more fiducial elements~\cite{Nielsen_2021}.
We also remark that assuming that $\rho_0$ and $M_{0}$ can be repeatedly accessed in an  experiment is by no means equivalent to having prior information about their representation, which could actually be any $\rho_0\in\mathsf{D}(\mathcal{H}),M_0\in\mathsf{C}(\mathcal{H})$. In fact, GST provides a representation not only for each gate but also for the native elements $\rho_0$ and $M_0$, which leads to  the gate set $\mathcal{G}$ in Eq.~\eqref{eq:gate_set}.


Rather than focusing on a  single process as in QPT~\eqref{eq:MLE_QPT}, the measured probabilities in GST account for the action of the three consecutive circuit parts of  Fig. \ref{fig: tomography_scheme}\textcolor{magenta}{c}, such that
\begin{equation}
\label{eq: probs_Appendix_GST}
p_{b, m_b, i, s}=   \mathrm{Tr}\left\{M_{z,m_b}\mathcal{E}_{b}\circ\mathcal{E}_i\circ\mathcal{E}_s(\rho_0)\!\right\}=\langle\!\langle M_{z,m_b}|H_{b}G_iH_{s}|\rho_0\rangle\!\rangle,
\end{equation}
where we have defined the imperfect binary fiducial POVM elements  as $M_{z,+}=M_0$ and $M_{z,-}=\mathbb{1}_2-M_0$. When using the process matrix representation, we keep the symbol $\mathcal{E}_i$ for the quantum channel denoting that gate, and introduce $\mathcal{E}_s$ and $\mathcal{E}_b$ for the imperfect operations that would transform the fiducial operators into the IC set in the ideal error-free situation.

Paralleling the discussion of QST and QPT, GST can proceed by either  inversion or by constrained optimization. When opting for the former, one obtains the linear GST estimates 
\begin{equation}
\label{eq:linear_GST}
\begin{split}
|\hat{\rho}_0\rangle\!\rangle = Tg|R_0\rangle\!\rangle,\hspace{1.5ex} \hat{G}_i=Tg^{-1}F_iT^{-1},\hspace{1.5ex}
\langle\!\langle \hat{M}_{0}|=\langle\!\langle Q_0|T^{-1}.
\end{split}
\end{equation}
Here, we have introduced the so-called {Gram matrix} $[g]_{\mu,s}=\langle\!\langle M_{\mu}|\rho_s\rangle\!\rangle=\langle\!\langle M_0|G_{\mu}G_s|\rho_0\rangle\!\rangle$ that contains information about the imperfect SPAM. Similarly, $[R_0]_{\mu}=\langle\!\langle M_{\mu}|\rho_0\rangle\!\rangle$ and $[Q_0]_s=\langle\!\langle M_0|\rho_s\rangle\!\rangle$ contain information about the imperfect measurement and imperfect state preparation, respectively. In the above expressions,  the  matrices $[F_i]_{\mu,s}=f_{\mu i s}$ contain the measured relative frequencies for each element of the gate set. A characteristic feature of GST is the appearance  of the matrix $T$, which can actually correspond to any invertible matrix, and cannot be fixed by looking solely at the measured relative frequencies. This means that GST estimates a gate set  up to a similarity transformation $T$, which acts simultaneously on the initial state, gate sets, and POVM elements  according to the equation above~\cite{Nielsen_2021}. This ambiguity, also referred to as {gauge redundancy} in the literature, makes experimental data compatible with (infinitely) many descriptions of the gate set, and  is not exclusive of GST \cite{Proctor_2017}. Just as occurs for gauge theories describing the most fundamental interactions of nature, there is a redundancy in our description of the GST experiments involving the sequential stages of faulty initialization, application of imperfect circuits, and faulty measurements, which will always lead to observed  quantities that are independent of the ``gauge", i.e. the choice of the $T$ matrix.

Let us note that the parallelism is not perfect in the sense that the above similarity transformation cannot happen `locally' at a specific location of the circuit, but must instead always occur globally. Additionally, there is no notion of gauge fields as the fundamental degrees of redundancy and, more importantly, this redundancy is not built on what is believed to be a fundamental symmetry of nature, but rests instead in the type of GST experiments that one performs. There can be other completely different experiments where this redundancy does not appear, which ultimately rest on our microscopic theories of light-matter interaction that underlie the implementation of the gates in the QIP device, but actually go much beyond the computational description of this device as a collection of qubits that can be subject to unitaries, or faulty versions thereof. The fact that we do not have access to the  `true' set of gates that is physically implemented in the  experiment, but rather to any gauge-equivalent set that results from the GST estimation is not problematic in general. However, it makes the  computation of distance metrics to a target gate set uncertain, as those metrics typically depend on the  gauge choice. To circumvent this issue, gauge optimization techniques are typically applied~\cite{Nielsen_2021}, establishing a privileged reference frame which allows to fix the similarity transformation $T$. 

Similarly to our discussion of linear-inversion QPT~\eqref{eq: linear_QPT}, non-physical estimates $\hat{\mathcal{G}}$ of the full  gate set $\mathcal{G}$ can also result from the linear GST~\eqref{eq:linear_GST}. Consequently, it is again advantageous to employ MLE techniques, albeit bearing in  mind that one will have to face  additional complications. First of all, the non-linearity of the probabilities on the estimation parameters (\ref{eq: probs_Appendix_GST}) makes the optimization problem no longer convex. Coupled with the above gauge redundancy, this conducts to the appearance  of local and degenerate minima in the landscape of the cost function, which complicates considerably the task of finding accurate estimates, and underscores the importance of selecting a sufficiently-accurate initial guess of the gate set. Secondly, and perhaps more importantly in the long run, GST scales exponentially in the number of qubits. It is important to recall that a quantum channel possesses $N_p=2^{2N}(2^{2N}-1)$ real parameters, being $N$ the number of qubits of the processor. This parameter count is in addition to those used to describe the native SPAM elements $N_{SPAM}$ and is further scaled by the number of gates that one aims to characterize, denoted as $N_G$. The total number of parameters is thus $N_{GST}=N_{SPAM}+N_G\times4^{N}(4^{N}-1)$, which  grows very fast and becomes intractable even for small qubit numbers.

Finally, it is noteworthy that the precision with which a gate set is estimated can be enhanced even with a limited number of experiments \cite{Nielsen2021}. Long-sequence gate set tomography (GST) addresses this by distributing shots across deeper circuits, which are constructed by repeating certain gates sandwiched between the SPAM elements $p$ times
\begin{equation}
p_{b,m_b, i, s}= \langle\!\langle M_{z,m_b}|H_b(G_i)^pH_s|\rho_0\rangle\!\rangle.
\end{equation}
This can result in an amplification  of the sensitivity of the measurements to a certain subset of parameters of the gate set, which underlies the reduction of the estimate error by a factor of $1/p$. In practice, individual gates are usually insufficient to amplify errors and achieving this effect \cite{Nielsen2021, ostrove2023nearminimal}, so combinations of them  should be employed instead, which are referred to as {germs} in the GST parlance.  The expression of the GST probabilities (\ref{eq: probs_Appendix_GST}) in terms of the corresponding process matrices  is shown in Equation (\ref{eq: probabilities}) of the main text, where we allow for circuits composed of several gates drawn from the gate set for long-sequence GST.  We refer to Sec.~\ref{sec_1} for a detailed discussion of the main long-sequence GST aspects, as well as to reference \cite{Nielsen_2021}. Interesting problems in this regard include {germ selection} and {fiducial pair reduction}.

\section{Coloured phase noise  for single-qubit gates}
\label{appendix_1}
In this Appendix, we describe the central aspects of the phase-noise parametrization of noisy gates employed in the main text, which builds on the recent results for the quantum dynamical map of trapped-ion gates presented in~\cite{velazquez2024dynamical}. In order to parametrize the full gate set in Eq.~\eqref{eq: gate_set}, we need to generalise this  quantum dynamical map to other driving phases, allowing for  the SPAM operations required for  IC.

Following~\cite{velazquez2024dynamical},  we start by modelling phase noise by   a stochastic process $\tilde{\delta}(t)$. For single-qubit gates, the microscopic rotating-frame Hamiltonian for a generic phase  reads
\begin{equation}
\begin{split}
\label{eq:stoch_H_Appendix}
	\tilde{H}(t) &= \frac{\Omega}{2}  \big(\cos\phi\sigma_x-\sin\phi\sigma_y\big)+ \frac{\tilde{\delta}(t)}{2} \sigma_z,
\end{split}
\end{equation}
where $\Omega$ $(\phi)$ is the  Rabi frequency (phase) of the drive, and 
\beq
  \tilde{\delta}(t)=\delta\tilde{\omega}(t)+\frac{{\rm d}}{{\rm d}t}\delta \tilde{\phi}(t)
\eeq
includes the frequency (phase) fluctuations $\delta\tilde{\omega}(t)$ $(\delta \tilde{\phi}(t))$  described by a specific stochastic process.
As a result, the quantum states become stochastic $\tilde{\rho}(t)=\ket{\tilde{\psi}(t)}\!\!\bra{\tilde{\psi}(t)}$, and its Liouville–von Neumann  equation which corresponds to a system of {stochastic differential equations} with multiplicative noise. One can formally derive the  equations  of motion for the statistical average ${\rho}(t)=\mathbb{E}[\tilde{\rho}(t)]$ by using
the   Nakajima-Zwanzig approach~\cite{10.1143/PTP.20.948,Zwanzig1960EnsembleMI,Breuer2002} of projection operators. As discussed in~\cite{velazquez2024dynamical}, to get a tractable and useful expansion for high-fidelity gates, one utilizes the so-called time-convolutionless methods~\cite{Chaturvedi1979,BRE02} in the {instantaneous or dressed-state basis} \cite{PAvan_1977,Biercuk_2011, Green_2013}. By truncating at second order, and setting $\phi=0$,  one arrives at a simple time-local master equation 
that can be expressed in terms of integrals of the covariance 
\beq
\label{eq:autocorrrelation_function}
C(t,t')=\mathbb{E}[ \tilde{\delta}(t) \tilde{\delta}(t')],
\eeq
where we assume that $\mathbb{E}[ \tilde{\delta}(t)]=0$. In the dressed-state basis  $\ket{\pm} = (\ket{0}\pm\ket{1})/\sqrt{2}$, the evolution for the populations reads 

\beq
\label{eq:analytical_exact}
    \rho_{\pm\pm}(t)=\half\pm\half\big(2\rho_{++}(0)-1\big)\ee^{-\Gamma_1(t)}\!,
\eeq
where we have introduced the following decay function
\beq
\label{eq:gamma_1}
    \Gamma_1(t) = \!\!\int_0^t\!\!\diff t'\,\gamma_1(t'),\hspace{1ex} \gamma_1(t')=-\!\!\int_0^{t'}\!\!\!\diff t''C(t'-t'') \cos(\Omega(t'-t'')).
\eeq
 This is precisely one of the four parameters which we aim to estimate under GST (see Eq.~\ref{eq:Gammas} of the main text). The remaining parameters stem from the evolution of the dressed-state coherences of the density matrix~\cite{velazquez2024dynamical}. This evolution is approximated by  a Magnus expansion~\cite{BLANES2009151} to lowest order, which yields the following integrals

\begin{equation}
\label{eq:gammas_mus}
    \Gamma_n(t) = \int_0^t\diff t' \gamma_n(t'), \quad \Delta_n(t) = \int_0^t \diff t'\delta_n(t'),
\end{equation}
 where we have introduced the remaining integrals
\beq
\label{eq:mus}
\begin{split}
    \gamma_2(t')&=-\int_0^{t'}\!\!\diff t''\,C(t'-t'') \cos(\Omega(t'+t'')),\\
    \delta_1(t')&=-\int_0^{t'}\!\!\diff t''\,C(t'-t'') \sin(\Omega(t'-t'')),\\
    \delta_2(t')&=+\int_0^{t'}\!\!\diff t''\,C(t'-t'') \sin(\Omega(t'+t'')).
\end{split}
\eeq
These parameters contain all the information that is required in our parametrized GST (see Sec.~\ref{mic-parameterization}), where the following combination shall be useful for the parametrization
\beq
\label{eq: module_Theta}
 \Theta(t) = \sqrt{\Delta_1^2(t)-\Delta_2^2(t)-\Gamma_2^2(t)}.
 \eeq

We can rewrite these parameters  in terms of the {power spectral density} (PSD) of the noise which, when restricting to wide-sense stationary processes $C(t,t')=C(t-t')$~\cite{solo1992intrinsic},  is defined  by the Fourier transform
\beq
\label{eq:psd_def}
C(t)=\int_{-\infty}^{\infty}\frac{{\rm d}\omega}{2\pi} S(\omega)\ee^{\ii\omega t}.
\eeq
This enables one to rewrite the integrals Eq.~(\ref{eq:gammas_mus}) as
\beq
\label{eq:Gammas_app}
\begin{split}
	\Gamma_n(t) = & \int_{-\infty}^\infty \!\!{\diff\omega}\,S(\omega)\,F_{\Gamma_n}(\omega,\Omega,t),\\
	\Delta_n(t) = & \int_{-\infty}^\infty \!\!{\diff\omega}\,S(\omega)\,F_{\Delta_n}(\omega,\Omega,t),
\end{split}
\eeq
where $F_{\Gamma_i}$ and $F_{\Delta_i}$ correspond to the filter functions mentioned during the introduction. In particular, one finds 
\beq
\label{eq:Gamma1_filter}
	F_{\,\Gamma_1}(\omega,\Omega,t) = \frac{t}{4}\!\left(\eta_{\frac{2}{t}\!}(\Omega-\omega)+\eta_{\frac{2}{t}\!}(\Omega+\omega)\right),\\
 \eeq
where we have used the nascent Dirac delta
\beq
\eta_{\epsilon}(x)=\frac{\epsilon}{\pi x^2}\sin^2\left(\frac{x}{\epsilon}\right)
\eeq
fulfilling $\eta_\epsilon(x)\to\delta(x)$ as $\epsilon\to0^+$, and $\int_{-\infty}^{\infty}{\rm d}x\,\eta_\epsilon(x)=1$, which clearly acts as a Dirac-type delta filter in the long-time limit. The other filters are defined as follows
\beq
 \label{eq:M1_filter}
  F_{\,\Delta_1}(\omega,\Omega,t) = \frac{\Omega t}{2\pi(\Omega^2-\omega^2)}+\delta F_{\,\Delta_1}(\omega,\Omega,t),
\eeq
where we have introduced 
\beq
	\delta F_{\,\Delta_1}(\omega,\Omega,t) = \frac{1}{4\pi}\left(\frac{\sin\big((\omega-\Omega)t\big)}{(\omega-\Omega)^2}-\frac{\sin\big((\omega+\Omega)t\big)}{(\omega+\Omega)^2}\right).
\eeq
In addition,  one also finds
\beq
\begin{split}
\label{eq:noN_{t}arkov_filter_functions}
	&F_{\,\Gamma_2}(\omega,\Omega,t) = \frac{2 \cos\Omega t}{\pi(\omega^2-\Omega^2)}\sin(\half(\omega-\Omega) t)\sin(\half(\omega+\Omega) t),\\
	&F_{\,\Delta_2} (\omega,\Omega,t)= \frac{2 \sin\Omega t}{\pi(\omega^2-\Omega^2)}\sin(\half(\omega-\Omega) t)\sin(\half(\omega+\Omega) t).
\end{split}
\eeq
These expression are of particular relevance because the phase-noise PSD can often be accessed experimentally independently of the tomographic characterization, e.g. by a self-heterodyne setup with a reference ultra-stable oscillator in trapped-ion devices~\cite{freund2023selfreferenced}. This would enable to adapt our GST to a specific QIP. For instance, as mentioned in Sec.~\ref{sec_1}, the validity of long-sequence GST is limited to the region in which the sequence depth fulfills $p\ll 1/\Gamma_1$. Thus, knowing the experimental PSD allows us to give a prior estimation of this region  by means of Eq.~(\ref{eq:Gammas}).

The former expressions also allow us to explicitly compute the non-Markovianity measure~\cite{PhysRevLett.105.050403,PhysRevA.89.042120} employed in Sec.~\ref{sec_2}. As a notion of quantum non-Markovianity, we use the so-called CP-divisibility
\beq
\label{eq:CP_div}
\mathcal{E}_{t,t_0}=\mathcal{E}_{t,t'}\circ\mathcal{E}_{t',t_0},\forall t'\in[t_0,t];\,\, \mathcal{E}_{t,t'}\in\mathsf{C}(\mathcal{H}_{\rm S}).
\eeq

Within this framework, a map is said to be Markovian if one can divide it by the composition of two CPTP maps at each intermediate time.
Therefore, acting on the tensor product space $\mathcal{H}_{\rm S}\otimes\mathcal{H}_{\rm S}$, one can build a non-Markovianity measure on the violation of the CPTP condition for each interval of evolution
\beq
g(t')=\lim_{\Delta t\to 0}\frac{1}{\Delta t}\bigg(\big\lVert\mathcal{E}_{t'+\Delta t,t'}\otimes\mathbb{1}(\ket{\Phi_+}\!\bra{\Phi_+})\big\rVert_1-1\bigg),
\eeq
where the trace norm is used, $\ket{\Phi_+}=\textstyle{\frac{1}{\sqrt{2}}}(\ket{00}+\ket{11})$ and the non-Markovianity measure is defined as $\mathcal{N}_{\rm CP}=\int_0^t{\rm d}t'g(t')$. Explicitly, using the integrals in (\ref{eq:gamma_1}) and (\ref{eq:mus}), one gets~\cite{velazquez2024dynamical}
\beq
\label{eq:NM_measure}
\mathcal{N}_{\rm CP}(t)=\frac{1}{2}\!\int_0^t\!\!{\rm d}t' \big(|\bar{\gamma}_-(t')|-\bar{\gamma}_-(t')\big),
\eeq
where $\bar{\gamma}_-(t)=\half\gamma_1(t)-\half\sqrt{\gamma_2^2(t)+\delta_2^2(t)}$.


Finally, one can use the analytic expressions for the time evolution of the dressed-state density matrix to derive closed expressions for the $\chi_{\phi}(t)$  matrix of dynamical quantum map. In the Pauli basis, this reads
\begin{equation}
\label{eq:chi_app}
\chi_0(t) = \frac{1}{2}\left( 
\begin{matrix} 
  \chi_A^{(1,1)}(t) & \chi_A^{(1,2)}(t) & 0 & 0 \\ 
   \chi_A^{(2,1)}(t) & \chi_A^{(2,2)}(t) & 0 & 0 \\ 
  0 & 0 & \chi_B^{(1,1)}(t) & \chi_B^{(1,2)}(t)\\ 
  0 & 0 & \chi_B^{(2,1)}(t) & \chi_B^{(2,2)}(t)\\
  \end{matrix}
  \right)
\end{equation}
where the upper and lower blocks can be expressed in terms of the parameters and the  Pauli matrices as follows
\begin{widetext}
\beq\begin{split}
\label{eq:chi_non_markov_A}
    \chi_A(t) &= \!\!\left(1+\ee^{-\Gamma_1(t)}\!\right)\!\!\mathbb{1}_2
    +2 \ee^{-\frac{1}{2}\Gamma_1(t)}\left[\!\bigl(\cos\Omega t\cos\half\Theta(t)-\frac{\Delta_1(t)}{\Theta(t)}\sin\Omega t\sin\half\Theta(t)\bigr)\!\sigma_z
    -\!\!\bigl(\sin\Omega t\cos\half\Theta(t)+\frac{\Delta_1(t)}{\Theta(t)}\cos\Omega t\sin\half\Theta(t)\bigr)\!\sigma_y\right],\\
    \chi_B(t)&= \!\!\left(1-\ee^{-\Gamma_1(t)}\!\right)\!\!\mathbb{1}_2
    -\frac{2}{\Theta(t)}\ee^{-\frac{1}{2}\Gamma_1(t)}\sin\half\Theta(t)\bigl[\bigl(\Gamma_2(t)\cos\Omega t+\Delta_2(t) \sin\Omega t\bigr)\sigma_z
    +\bigl(\Gamma_2(t)\sin\Omega t-\Delta_2(t)\cos\Omega t\bigr)\sigma_x\bigr].
\end{split}\eeq
\end{widetext}
Note that nn contrast to~\cite{velazquez2024dynamical}, we here present the full $\chi_{\phi}(t)$  matrix and not only that of the error channel. We also note that, in the absence of noise $\Gamma_i(t)=\Delta_i(t)=\Theta(t)=0$, the process block matrices are simply  
\beq
\chi_A(t)=2(\mathbb{1}_2+\cos\Omega t\sigma_z)-2\ii\sin\Omega t\sigma_y,\hspace{1ex} \chi_B(t)=0,
\eeq
and one recovers the perfect gate $\mathcal{E}(\rho)=\ee^{-\ii t\frac{\Omega}{2}\sigma_x}\rho\ee^{+\ii t\frac{\Omega}{2}\sigma_x}$.

For the present work, we need to generalise this derivation   for different phase angles $\phi$ in order to attain IC. We remark that the process matrices accounting for the same pulse duration are paramterized by the same filter integrals $\Gamma_i(t)$ and $\Delta_i(t)$, irrespectively of their phase. In fact, for the gate set used in this work, the $G_{\pi}$ matrix is obtained just by adding a minus sign to the non-diagonal elements of Eq.~(\ref{eq:chi_app}). For $G_{\pi/2}$ we find

\begin{equation}
\label{eq:chi_app_pi_2}
\chi_{\pi/2}(t) = \frac{1}{2}\left( 
\begin{matrix} 
  \chi_A^{(1,1)}(t) & 0 & -\chi_A^{(1,2)}(t) & 0 \\ 
   0 & \chi_B^{(1,1)}(t) & 0 & \chi_B^{(1,2)}(t) \\ 
  -\chi_A^{(2,1)}(t) & 0 & \chi_A^{(2,2)}(t) & 0\\ 
  0 & \chi_B^{(2,1)}(t) & 0 & \chi_B^{(2,2)}(t)
\end{matrix} 
\right),
\end{equation}
and finally for $G_{3\pi/2}$
\begin{equation}
\label{eq:chi_app_3pi_2}
\chi_{3\pi/2}(t) = \frac{1}{2}\left( 
\begin{matrix} 
  \chi_A^{(1,1)}(t) & 0 & \chi_A^{(1,2)}(t) & 0 \\ 
   0 & \chi_B^{(1,1)}(t) & 0 & -\chi_B^{(1,2)}(t) \\ 
  \chi_A^{(2,1)}(t) & 0 & \chi_A^{(2,2)}(t) & 0\\ 
  0 & -\chi_B^{(2,1)}(t) & 0 & \chi_B^{(2,2)}(t)
\end{matrix} 
\right).
\end{equation}

Let us finally consider additional amplitude noise fluctuations, as mentioned in Subsec.~\ref{amplitude noise}. As exposed in \cite{velazquez2024dynamical}, the drive amplitude fluctuations will be modeled by stochastic noise in the Rabi frequency~\cite{Zoller_1978} roughly as
\begin{equation}
\label{eq: rabi_noise}    \Omega\to\Omega+\delta\tilde{\Omega}(t),\hspace{1ex} \delta\tilde{\Omega}(t)\simeq \frac{\Omega}{2}\frac{\delta \tilde{I}_L(t)}{I_{L,0}},
\end{equation}
where $\delta\tilde{\Omega}(t)$ and $\delta \tilde{I}_L(t)$ denote the stochastic fluctuations of the Rabi rate and the laser intensity, respectively, and $I_{L,0}$ denotes the ideal intensity of the driving field.

Note that the Hamiltonian Eq.~(\ref{eq:stoch_H_Appendix}) should now include stochastic noise in the Rabi frequency ($\Omega\to\Omega+\delta\tilde{\Omega}(t)$). For the ongoing analysis, we assume that the stochastic processes describing phase and amplitude fluctuations are not cross-correlated. That is, $\mathbb{E}\{\tilde{\delta}(t)\delta\tilde{\Omega}(t')\}=0$. Under an equivalent procedure to the that exposed for pure phase noise gates in this appendix, one arrives to expressions for the evolution of the states which are described by the filtered integrals Eq.~(\ref{eq:Gammas_app}) supplemented with a correction to the decay $\Gamma_1$
\begin{equation}\label{eq: amplitude_correction_apemdix}
    \Delta\Gamma_1(t) = \int_{-\infty}^\infty \!\!\!{\rm d}\omega {S}_{\Omega}(\omega)F_\Omega(\omega,t),
\end{equation}
where ${S}_{\Omega}$ stands for the PSD of the amplitude noise and we have introduced an additional decay filter function 
\begin{equation}
    \label{eq:filterDeltaGamma}
    F_\Omega(\omega,t)= \frac{1}{\pi} \left(\frac{1-\cos(\omega t)}{\omega^2}\right)=t\eta_{\frac{2
}{t}}(\omega).
\end{equation}

We note here that the inclusion of this new parameter does not influence the structure of the matrices Eqs.~\eqref{eq:chi_app}, \eqref{eq:chi_app_pi_2} and ~\eqref{eq:chi_app_3pi_2}, which is preserved. However, the upper and lower blocks Eq.~(\ref{eq:chi_non_markov_A}) must be updated to account for this correction

\begin{widetext}
\beq\begin{split}
\label{eq:chi_non_markov_A_amplitude}
    \chi_A(t) &= \!\!\left(1+\ee^{-\Gamma_1(t)}\!\right)\!\!\mathbb{1}_2
    +2 \ee^{-\frac{1}{2}(\Gamma_1(t)+\Delta\Gamma_1)}\left[\!\bigl(\cos\Omega t\cos\half\Theta(t)-\frac{\Delta_1(t)}{\Theta(t)}\sin\Omega t\sin\half\Theta(t)\bigr)\!\sigma_z\right.-\\
    &\hspace{60ex}-\left.\!\!\bigl(\sin\Omega t\cos\half\Theta(t)+\frac{\Delta_1(t)}{\Theta(t)}\cos\Omega t\sin\half\Theta(t)\bigr)\!\sigma_y\right],\\
    \chi_B(t)&= \!\!\left(1-\ee^{-\Gamma_1(t)}\!\right)\!\!\mathbb{1}_2
    -\frac{2}{\Theta(t)}\ee^{-\frac{1}{2}(\Gamma_1(t)+\Delta\Gamma_1)}\sin\half\Theta(t)\bigl[\bigl(\Gamma_2(t)\cos\Omega t+\Delta_2(t) \sin\Omega t\bigr)\sigma_z
    +\bigl(\Gamma_2(t)\sin\Omega t-\Delta_2(t)\cos\Omega t\bigr)\sigma_x\bigr].
\end{split}\eeq
\end{widetext}



\bibliographystyle{apsrev4-1}
\bibliography{biblio}

\end{document}